\documentclass [12pt]{article}
\usepackage{latexsym}
\usepackage{amsfonts}
\usepackage{acronym}
\newcommand{\be}{\begin{equation}}
\newcommand{\ee}{\end{equation}}
\newcommand{\bea}{\begin{eqnarray}}
\newcommand{\eea}{\end{eqnarray}}

\begin{document}
\normalsize
\title{\Large Dirac fundamental quantization of gauge theories is natural way of reference frames in modern physics.}
\author
{{\bf L.~D.~Lantsman}\\
 Wissenschaftliche Gesellschaft bei
 J$\ddot u$dische Gemeinde  zu Rostock,\\Augusten Strasse, 20,\\
 18055, Rostock, Germany; \\ 
Tel.  049-0381-799-07-24,\\
llantsman@freenet.de}
\maketitle
\begin {abstract}
We analyse two principal approaches to the quantization of physical models. 
There are the Faddeev-Popov "heuristic" approach, based on fixing a gauge in the FP path integrals formalism, and the "fundamental" approach of Dirac based on the constraint-shell  reduction of Hamiltonians with  deleting of unphysical variables. 
The relativistically invariant FP "heuristic" approach deals with the enough small class of problems associated with S-matrices squared considering on-shell  quantum fields.
On the other hand, the "fundamental" quantization approach of Dirac involves the manifest relativistic covariance of quantum fields that survives the constraint-shell  reduction of Hamiltonians. 
One can  apply this approach to a broader class of problems than by studying  S-matrices.  Investigations of various bound states in QED and QCD are patterns of such applications.
In the present  study, with the example of the Dirac "fundamental" quantization of the Minkowskian non-Abelian Higgs model (studied in its historical retrospective), we  show  obvious advantages of this quantization approach. The arguments in favour of the Dirac fundamental
quantization of a physical model will be presented as a way of Einstein and Galilei
relativity in modern physics.

\end{abstract}
\section {Introduction}
 The modern gauge physics developed in such a way that the  quantization approach by Feynman \cite{Feynman1}, referring to as the {\it heuristic} one, became the main to the end of 60-ies. 
Calculating radiation corrections to scattering processes, Feynman has elucidated that scattering amplitudes of elementary particles in the perturbation theory  does not depend on a reference frame and choice of the gauge \footnote{Indeed, as we shall discuss below, repeating the arguments \cite {Bogolubov-Shirkov}, only scattering amplitudes squared are relativistically invariant.}.
Utilizing this fact, it is possible to alter the QED Lagrangian turning it in a gauge model without constraints. 
The independence of a reference frame (we shall refer to is as  the 
S-invariance in the present study), recently also named {\it the relativistic invariance}, while {\it gauge fixing } came to a formal procedure of choice of gauge covariant field variables.

\medskip
Indeed, an imperceptible substitution of the sense of notions in the method of gauge (G) covariant and S-invariant heuristic quantization \cite{Feynman1} has occurred. An
alternative approach to the quantization of gauge (non-Abelian) theories is known
 \cite{Feynman1} about the quantization of gauge (non-Abelian) theories.
In this approach of Dirac, finding the S-covariant and G-invariant solutions to {\it constraint equations} was proposed.
The name {\it fundamental quantization} to  this quantization method was devised by  Schwinger   \cite{Schwinger}.

Briefly, the strategy of the  fundamental quantization approach \cite{Dir}  was the following.

1. One would utilize the constraint equations and G-invariance in order to remove  unphysical variables (degrees of freedom) and construct  G-invariant nonlocal functionals of gauge fields, so-called {\it Dirac variables} \cite{Dir}.  
  In particular,  there was demonstrated in Ref. \cite{Dir} by utilizing Dirac variables that solving the QED equations in the class of mentioned nonlocal functionals of gauge fields involves the Coulomb (radiative) gauge for electromagnetic fields. 

2. One would also prove  the S-covariance on the level of Poincar$\grave{}e$ generators for G-invariant observables.  
 One of the first proofs is due  to  Zumino \cite{Zumino}.
The dependence of G-invariant observables on the chosen reference frame parameters is called the {\it implicit relativistic covariance}.

3. Finally, one would construct the S-covariant S-matrix in terms of G-invariant observables.

\medskip
 This program regarding QED was stated in the review \cite{Polubarinov}.
One can discover that series of well-known facts and conclusions of QED was interpreted therein not as it is customary in modern literature. 
For instance, the Coulomb field is the precise consequence of solving the one of classical  equations ({\it the Gauss law equation}) but on no account of the large mass approximation. 
Herewith the action functional of QED taking in the Coulomb gauge is the 
"one-to-one" consequence of solving the Gauss law equation in terms of 
G-invariant Dirac variables \cite{Dir} and not (only) the result of  choice  a gauge.
As an example, a proton and electron in a relativistic moved atom form this atom due to the Coulomb field transformed into the appropriate Lorentz reference frame and not as a result of  an interaction described by additional Feynman diagrams.

\medskip
 After  this brief analysis of the fundamental quantization approach \cite{Dir,Polubarinov} it becomes obvious complete  substitution this approach by the "heuristic" \cite{Feynman1} one. 
It would be  necessary to prove the relativistic covariance on the level of Poincare generators for G-invariant observables, if the result of  computations for scattering amplitudes is S-invariant, i.e. does not depend on a reference frame.
Also the  question is what G-invariant observables are necessary if one can utilize various variables, including those for solving problems constructing the unitary perturbation theory and proving the renormalizibility of the Standard model.
Formulating and solving these actual problems implemented in the framework of the heuristic quantization led to the situation when  this quantization approach became, in fact, the only method with which one associated solving of problems in the modern field theory.
One forgets,  however, that the application sphere of the heuristic quantization is restricted strictly to the problems of scattering of elementary particles (quantum fields)  where this quantization method has arisen \cite{Feynman1}.
For the needs of study of the bound-states physics, hadronisation and confinement, in describing the quantum universe, the fundamental quantization \cite{Dir,Polubarinov} is more adequate, as  Schwinger \cite{Schwinger} has predicted.

\medskip
 The present study is an attempt to compare in detail the both quantization methods: the fundamental and heuristic ones, with regard to the important sphere of modern theoretical physics, the non-Abelian gauge theory (although some aspects of QED, the typical Abelian gauge theory, will be also the subject of our discussion).

The present article is organized as follows.
In Section 2 we discuss in detail the fundamental and heuristic approaches to quantization of gauge theories.
Herewith the Faddeev-Popov (FP) "heuristic" quantization method \cite{FP1}, involving the FP path integrals formalism, as the modern realization of the Feynman approach  \cite{Feynman1} in the sphere of gauge physics, will be  investigated.
The principal result of Section 2 will be the demonstration that the  Feynman rules $({\rm FR})^F$ got in the FP path integrals formalism \cite{FP1} for a gauge model (when a gauge $F$ is fixed) coincide with the  Feynman rules $({\rm FR})^*$  got in the fundamental quantization formalism \cite{Dir} only for quantum fields on-shell described correctly by S-matrices. 
The latter statement may be treated as the {\it gauge equivalence} (or independence) theorem \cite{Taylor, Slavnov1}.  
On the other  hand, because of the manifest relativistic covariance of Green functions in  gauge models quantized by Dirac \cite{Dir}, in which the constraint-shell reduction of appropriate Hamiltonians is performed,  various {\it spurious} Feynman diagrams (SD) \cite{Pervush2,Nguen2}
appear in those models. 
As a result, on the level of the heuristic FP quantization \cite{FP1}, the appearance of spurious Feynman diagrams in constraint-shell gauge theories implies, for the on-shell of quantum fields,  the modification of the {\it gauge equivalence}  theorem \cite{Taylor, Slavnov1} in such a way that the  Feynman rules for SD would be added to the  Feynman rules $({\rm FR})^F$ for relativistic covariant Green functions.
However, when asymptotical states contain composite fields (say, hadronic bound states off-shell) or collective (vacuum) excitations, the {\it gauge equivalence}  theorem \cite{Taylor, Slavnov1} between the FP path integrals formalism \cite{FP1} and Dirac fundamental quantization method \cite{Dir} becomes  very problematic, and one  can be sure only in the above adding of the  Feynman rules for SD when such states are in question.
Violating the gauge equivalence theorem \cite{Taylor, Slavnov1} in this case does not mean the gauge non-invariance and relativistic  non-covariance. It reflects only the non-equivalence of the different definitions of sources in Feynman and FP path integrals because of nontrivial  boundary conditions and residual interactions forming asymptotical composites or collective states. \par 

\medskip
In Section 3,  with the example of the Dirac fundamental quantization \cite{Dir}
 of the Minkowskian non-Abelian Higgs model (studied in its historical retrospective), we demonstrate obvious advantages of this quantization approach in comparison with the Feynman-FP "heuristic" quantization method \cite{Feynman1, FP1}, when the topologically  nontrivial dynamics is taken into account.

\medskip
In Section 4 we discuss the future perspectives of development of the Minkowskian non-Abelian Higgs model quantized by Dirac.
It will be argued
 in favour of the "discrete" vacuum geometry as that justifying various effects associated with the Dirac fundamental quantization \cite{Dir} of that model.
\section{Comparison of heuristic and fundamental quantization schemes} 
The essence of the { heuristic} FP approach \cite{FP1} to quantization of gauge theories, logically continuing the Feynman method \cite{Feynman1}, is fixing  a gauge (say, $F(A)=0$) by the so-called {\it Faddeev trick}~: a gauge is fixed in an unique wise within an orbit of the appropriate gauge group (to within the {\it Gribov ambiguity} in specifying the transverse gauges \cite{Al.S., Gribov, Baal, Sobreiro} in non-Abelian gauge theories). 

\medskip
It will be now appropriately to recall some features of the FP heuristic quantization of non-Abelian gauge theories, the important part of modern gauge physics (QCD, the electroweak and Standard models).
The Gribov ambiguity in non-Abelian gauge theories considering in the transverse (Landau) gauge \cite{Sobreiro} $\partial_\mu A_\mu =0$  comes to FP path integrals regular (nonzero) out of the {\it Gribov horizon} $\partial \Omega$ \cite{Gribov, Baal, Sobreiro}. 
This horizon may be defined \cite{Gribov, Baal, Sobreiro} as the boundary  of the {\it Gribov region} (in the coordinate space) where the FP operator \cite{Sobreiro} 
\be \label{FP operator}
\Delta_{\rm FP} \equiv  \partial_\mu (\partial_\mu \cdot +[ A_\mu, \cdot])
\ee 
is nonnegative  \footnote{ In general \cite{Sobreiro}, there is the countable number of Gribov regions, $C_0$, $C_1,\dots$,
 in an (Euclidian) non-Abelian gauge theory where the Landau gauge $\partial_\mu A_\mu =0$ is not taken. Herewith subscript indices $0,1,\dots$ denote the numbers of zeros of the FP operator $\Delta_{\rm FP}$ in the appropriate Gribov region.

But with taking the transverse Landau gauge, only the one Gribov region, $C_0$, survives. The  FP operator $\Delta _{\rm FP} $ is positive inside this region, but attains its (infinitely degenerated) zero on its boundary, the { Gribov horizon} $\partial \Omega$.
Herewith it becomes evident that the Gribov horizon $\partial \Omega$ (in the coordinate representation) coincides with the light cone $p^2\equiv  -\partial_\mu \partial^\mu =0$. 

The said may serve as a (perhaps enough roughly, but obvious) description for the  Gribov ambiguity \cite{Al.S., Gribov, Baal, Sobreiro} in non-Abelian gauge theories.}.

Thus in non-Abelian gauge theories  in which the transverse Landau gauge $\partial_\mu A_\mu =0$ is fixed, the appropriate FP path integrals  become                                                                                                                                                              singular over the light cone \linebreak $p^2\equiv  -\partial_\mu \partial^\mu =0$ coinciding with the Gribov horizon $\partial \Omega$ \cite{Gribov, Baal, Sobreiro}. 

Finally, the (non-Abelian) FP path integrals for gauge models involving gauge fields $A$ and fermionic ones, $\psi $ and $\bar \psi $, acquire the look \cite{Pervush2}:
\begin{eqnarray}
\label{fpi11}  Z^{FP}[s^F, {\bar s}^F, J^F]\;=\;\int \prod_{\mu}DA^F_\mu D  \psi^F D{\bar \psi}^F \Delta _{FP}^F
 \delta (F(A^F)) e^{iW[ A^F,\psi^F {\bar \psi}^F ] + S^F},  
\end{eqnarray}  
with  $\Delta _{\rm FP}^F\equiv{\rm det}~ M_F $ being the  FP operator for the gauge $F(A)=0$ (in general, different from the transverse Landau gauge $\partial_\mu A_\mu =0$)  and
\be
\label{fpso1}  
S^F=\int d^4x \left( {\bar s}^F \psi^F +
 {\bar \psi}^F s^F  + A^F_{\mu}J^{\mu}\right)
 \ee
being  the sources term \footnote{ It will be also well-timed to cite here the explicit look of the \rm FP  determinant \rm $M_F$ in a (non-Abelian) gauge theory. It is \cite {Cheng}
\bea {\rm det} ~M_F &\sim &\int [d{\bf c}][d{\bf c}^\dagger] \exp \left \{ i \int d^4x d^4y \sum \limits_{a,b} c^\dagger_a (x)  ~M_F ^{ab}(x,y) c_b(y)\right \},\nonumber\eea
with ${\bf c}$ and ${\bf c}^\dagger $ being, respectively, FP ghost and 
anti-ghost fields. 

The FP  determinant \rm ${\rm det}~ M_F $ implies, for instance, the FP ghost action functional $S_{\rm FPG}$ \cite {Cheng},
$$ S_{\rm FPG}= \frac {1}{g} \int d^4x \sum \limits_{a,b} c^\dagger_a (x) \partial^\mu[\delta_{ab} \partial_\mu -g \epsilon_{abc} A_\mu^c ] c_b(x),$$
contributing obligatory to the total (non-Abelian) action as the Lorentz covariant gauge $\partial_\mu A^\mu$ is set.}.

Alternatively, the FP  operator \rm $M_F $ may be specified \cite {Cheng,LP1}  in terms of the linear response of the gauge $F(A)=0$ to a gauge transformation 
$$F (e^\Omega (A+\partial)e^{-\Omega}) =M_{F}\Omega +
O(\Omega^2)).$$

\medskip
The  approach \cite{FP1} to the heuristic quantization of gauge (non-Abelian)  theories had, of course, series of its unquestionable  services and successes: for instance, in constructing GUT, the universal model of gauge fields. 
 In particular, with the aid of the heuristic approach \cite{FP1}, the renormalizability of  GUT was proved.  \par
\medskip
But an essential shortcoming of the heuristic quantization method 
\cite{FP1} was "throwing off" of the notion "reference frame" from gauge physics. 
This notion is simply not necessary in that the method dealing with scattering  amplitudes of quantum fields \linebreak on-shell \footnote{ We recommend our readers \S 2 to Chapter 3 in the monograph \cite{Slavnov} where the FP integral for the "exact" YM theory, involving the manifest unbroken $SU(2)$ symmetry and only gauge fields, was derived utilising the properties of the appropriate $S$-matrix.
Indeed, the heuristic quantization approach \cite{FP1} involves the manifest relativistic invariance of \it local \rm scattering amplitudes squared, $\vert S_{fi}\vert ^2$, with $f$ and $i$ being, respectively, the final and initial states of colliding particles.  
However the scattering amplitudes $S$ are, indeed,  manifestly relativistically covariant (see e.g. \S 20.4 in \cite{Bogolubov-Shirkov}), and this  implies their manifest unitarity.  
On the other hand, probabilities of  scattering processes, that would be, doubtless,  relativistically invariant values, always involve scattering amplitudes squared.}. 
Thus FP path integrals induced by the "heuristic" quantization approach \cite{FP1} do not depend on anyone's choose of reference
frames. \par
\bigskip
It may be verified that in calculations of elements of S-matrices inherent in  gauge models on-shell the following obvious identity \cite{Arsen} for the appropriate
Feynman rules (FR)  takes place
\be \label{spur1}
({\rm FR})^F=({\rm FR})^*\quad ({\rm for~ S-matrices})
\ee
when the gauge $F$ is fixed. \par 
The expression $({\rm FR})^*$,  on the right-hand side of (\ref{spur1}), is referred to the Feynman rules in the considered gauge model upon performing the constraint-shell reduction of that model, involving ruling out of the unphysical (manifestly gauge covariant) field variables. 
This statement may be treated as the {\it gauge equivalence} (or independence) theorem \cite{Taylor, Slavnov1, Arsen}. \par
\medskip
But the diapason of problems solved in modern theoretical (in particular, gauge) physics is not restricted by the scattering processes of quantum fields 
on-shell.  
Among  such problems, one can point out the problem of (asymptotically)  bound and collective vacuum states.  These are patterns of composite quantum fields that are  off-shell of elementary particles. 
It turns out that the presence of such states in a quantum-field theory (QFT) may violate  the {gauge equivalence}  theorem \cite{Taylor, Slavnov1, Arsen}: at least it becomes quite problematic in this case. 
On the other hand, the constraint-shell (Hamiltonian) reduction of a gauge theory implies   ruling out of  the unphysical fields variables, i.e. describing this gauge theory in terms of only the  gauge invariant physical (observable) fields. 
In the so-called {\it particular} gauge theories (for instance, in the terminology \cite{Gitman}), examples of which are four-dimensional QED, the YM theory and QCD (i.e. Abelian as well as non-Abelian gauge models), involving the singular Hessian matrix  
\be \label{Hess}M_{ab}= \frac {\partial ^2 L}{\partial \dot q ^a\partial\dot q ^b}
\ee
(with $L$ being the Lagrangian of the considered gauge theory,   $q ^i$ being the appropriate degrees of
freedom and $\dot q^i$ being their time derivatives), the removal of unphysical degrees of freedom is associated, in the first place, with  ruling out  of the temporal components $A_0$ of gauge fields. 
In turn, it is associated with the zero canonical momenta $\partial  L/ \partial \dot q ^0$ conjugate to the fields $A_0$ in the particular gauge theories
$$ \partial  L/ \partial  \dot A_0 \equiv 0.$$
Thus temporal components $A_0$ of gauge fields are, indeed, non-dynamical degrees of freedom in particular theories, the quantization of which  contradicts the Heisenberg uncertainty principle. 
\par
Dirac \cite{Dir} and, after him,  other
authors of the first classical studies in quantization of gauge fields, for instance  
\cite{Heisenberg,Fermi},   eliminated temporal components of gauge
fields by gauge transformations. 
The typical look of such gauge transformations is \cite{David3}
\be \label{udalenie} v^T({\bf x},t)(A_0+\partial_0)(v^{T})^{-1}({\bf x},t)=0.\ee
 This equation may be treated as that specifying the gauge matrices $ v^T({\bf x},t)$.  
This, in turn, allows to write down the gauge transformations for spatial components of gauge
fields \cite{LP1}  (say, in a non-Abelian gauge theory) 
\be \label{per.Diraca} {\hat A}^D_i({\bf x},t):=v^T({\bf x},t)({\hat A}_i+\partial_i)(v^{T})^{-1}({\bf x},t); \quad {\hat A}_i= g \frac {\tau ^a}{2i}A_{ai}.
\ee
It is easy to check that the functionals ${\hat A}^D_i({\bf x},t)$ specified in such a way are gauge invariant and transverse fields  
\be \label{Dv} \partial_0  \partial_i {\hat A}^D_i({\bf x},t) =0; \quad  u ({\bf x},t)  {\hat A}^D_i({\bf x},t)   u ({\bf x},t)^{-1}= {\hat A}^D_i ({\bf x},t)    \ee 
for gauge matrices $ u ({\bf x},t)$. \par 
Following Dirac \cite{Dir}, we shall refer to  the  functionals ${\hat A}^D_i({\bf x},t)$ as to the {\it Dirac variables}.  
The Dirac variables ${\hat A}^D_i$ may be derived by resolving the Gauss law constraint
\be \label {Gau} \partial W/\partial A_0=0\ee
(with $W$ being the action functional of the considered gauge theory). \par 
 Solving Eq. (\ref{Gau}) \cite{Pervush2}, one expresses  temporal components $A_0$ of gauge
fields $A$ through their  spatial components; by  that the nondynamical components $A_0$ are indeed ruled out from the appropriate Hamiltonians.  
Thus the reduction of particular gauge theories occurs over the surfaces of the appropriate Gauss law constraints. 
Only upon expressing temporal components $A_0$ of gauge
fields $A$ through their  spatial components one can perform gauge transformations (\ref{per.Diraca}) in order to  turn spatial components $\hat A_i$ of gauge
fields into gauge-invariant and transverse Dirac variables ${\hat A}^D_i$ \cite{LP1}. Thus, formally, temporal components $A_0$ of these fields become zero.  
By that the Gauss law constraint (\ref{Gau}) acquires the form  \cite{Pervush2}
$$ \partial_0 \left(\partial_i \hat A^D_i({\bf x},t) \right) \equiv 0.$$
For  further detailed study of the "technology" getting Dirac variables, in particular gauge theories, we recommend  the works \cite{Pervush2, Nguen2, Azimov} (four-dimensional constraint-shell QED involving electronic currents) and \cite{LP1,LP2} (the Minkowskian non-Abelian Gauss law
constraint-shell model involving vacuum BPS monopole  solutions; we shall discuss it briefly also in the next section). \par
\medskip 
Dirac variables prove to be manifestly relativistically covariant.  
  Relativistic properties of Dirac variables in gauge theories were investigated in the papers \cite{Heisenberg} (with the reference to the unpublished note by von Neumann), and  then  this job was continued by I. V. Polubarinov in his review \cite{Polubarinov}. \par 
These investigations displayed that there exist such relativistic transformations of Dirac variables that maintain transverse gauges of  fields. 
More precisely,  Dirac variables ${\hat A}^{(0)D}$ observed in a rest  reference frame $\eta_\mu^0=(1,0,0,0)$ in the Minkowski space-time  (thus $\partial_i{\hat A}^{(0)D}_i=0$), in a moving reference frame 
\be \label{dvig} \eta ^{\prime}  = {\eta^0 }+\delta^0_L{\eta_0}    \ee
are also transverse, but now  regarding the new reference frame $\eta ^{\prime}$ \cite{Pervush2,David3} 
$$ \partial_\mu{\hat A}^{D\prime}_\mu=0.
$$
In particular, $A_0(\eta^0) = A_0(\eta^{0\prime}) =0$, i.e. the Dirac removal (\ref{udalenie}) \cite{Dir, David3} of temporal components of gauge fields is transferred from the rest to the moving reference frame. 
In this consideration \cite{Polubarinov, Pervush2,Heisenberg}, $\delta^0_L $  are ordinary total Lorentz transformations of coordinates, involving  appropriate transformations of fields (bosonic and fermionic).  
When one transforms fields entering the  gauge theory into Dirac variables \footnote{ It may be demonstrated \cite{Pervush2, Nguen2,Arsen,Azimov} that the  transformations (\ref{per.Diraca}), turning gauge fields $A$ into Dirac variables $\hat A^D$, imply the $\psi^D= v^T({\bf x},t)\psi$ transformations for fermionic fields $\psi$ and $\phi^D= v^T({\bf x},t)\phi$ transformations for spin 0 fields: to latter ones belong, for instance, Higgs vacuum BPS monopole solutions investigated in the recent papers \cite{LP1,LP2}.  } in a rest reference frame $\eta_0$ and then goes over to a moving reference frame $\eta^\prime$, Dirac variables $\hat A^D$, $\psi^D$, $\phi^D$   are suffered relativistic transformations consisting of two therms \cite{Pervush2, Nguen2}. \par 
The first therm is the response of Dirac variables on ordinary total Lorentz transformations of coordinates (Lorentz busts)
$$x'_k=  x_k+ \epsilon_ k t, \quad t'= t+ \epsilon_ k x_k, \quad \vert \epsilon_ k \vert \ll 1. $$
The second therm corresponds to "gauge" Lorentz transformations $\Lambda(x)$ of Dirac variables  $\hat A^D$, $\psi^D$, $\phi^D$   \cite{Pervush2, Nguen2}
$$  \Lambda(x)\sim \epsilon_ k \dot A^D_k (x)\Delta^{-1},    $$ 
with 
$$\frac{1}{\Delta}f(x)=-\frac{1}{4\pi}\int d^3y\frac{f(y)}{|\bf{x}-{\bf y}|}
$$
for any continuous function $ f(x)$.    
Thus any relativistic transformation for  Dirac variables may be represented as the sum of two enumerated therms. 
For instance \cite{Pervush2},
\begin{eqnarray}
\label{ltf} 
 A_{k}^{D} [ A_{i} &+& \delta_{L}^{0} A ] - A_{k}^{D} [ A ]  =  \delta_{L}^{0} A_{k}^{D} + \partial_{k} \Lambda,  \end{eqnarray} 
 \begin{eqnarray}
\label{ltf1}
 \psi^{D} [ A &+& \delta_{L}^{0} A , \psi + \delta_{L}^{0} \psi ] -  \psi^{D} [ A, \psi ] = \delta_{L}^{0} \psi^{D} + i e \Lambda  (x^{\prime}) \psi^{D}.  \end{eqnarray}  
 Relativistic transformations of  Dirac variables of the (\ref{ltf}), (\ref{ltf1}) type imply immediately definite relativistic transformations for Green functions inherent in constraint-shell (Gauss-shell) gauge theories.  
For example \cite{Nguen2}, in four-dimensional constraint-shell QED the electronic Green function
$$  G(p)=G_0(p)+ G_0(p) \Sigma (p) G_0(p) + O(\alpha^4), \quad  G_0(p)=[ p_\mu \gamma^\mu-m]^{-1},  $$ 
with $\Sigma (p) $ being the electronic self-energy, proves to be relativistic covariant under the "gauge" Lorentz transformations  $\Lambda(x)$.  
This, in turn, mathematically equivalent to the complete Lorentz invariance of  the electronic self-energy $\Sigma (p)$ \cite{Nguen2}
$$\delta_L^{\rm tot} \Sigma (p)= (\delta_L^0 +\delta_\Lambda) \Sigma (p)=0.
$$
\medskip
The relativistic covariance of  Green functions inherent in constraint-shell gauge theories implies the appearance of various {\it spurious} Feynman diagrams (SD) in those theories \cite{Pervush2,Nguen2}.   SD are generated \cite{Arsen} by gauge factors $v^T({\bf x},t)$. 
On the level of the heuristic FP quantization \cite{FP1}, the appearance of spurious Feynman diagrams in constraint-shell gauge theories implies,  on-shell of quantum fields,  the modification of the {gauge equivalence}  theorem \cite{Taylor, Slavnov1, Arsen} 
$$({\rm FR})^F+ ({\rm SD})\equiv ({\rm FR})^* \quad ({\rm for~ Green~ functions)}
$$
as a consequence of the independence of FP  path integrals (\ref{fpi}) on the choice of a reference frame.

When, however, asymptotical states contain composite fields (say, hadronic bound states off-shell) or collective (vacuum) excitations, the {gauge equivalence} theorem (\ref{spur1})  \cite{Taylor, Slavnov1, Arsen} becomes  problematic, and one  may be sure only in the identity 
\be \label{spur2} ({\rm FR})^F+ ({\rm SD})\equiv ({\rm FR})^* \quad ({\rm for~ S- matrices~ with ~composite~ fields}).
\ee
Violating the gauge equivalence theorem \cite{Taylor, Slavnov1, Arsen} in this case does not mean the gauge non-invariance and relativistic  non-covariance. It reflects only the non-equivalence of the different definitions of sources in Feynman and FP path integrals because of nontrivial  boundary conditions and residual interactions forming asymptotical composite or collective states. \par
More exactly, with a transverse gauge $F(A)=0$  fixed  (for instance, it is the Landau gauge $\partial_\mu  A_\mu=0$ in non-Abelian gauge models), the sources term $S^F$, (\ref{fpso1}), in the given  FP path integral  (\ref {fpi11}) is on-shell of quantum fields. 
In this case, in the fermionic sector of the considered gauge theory written down in terms of the FP path integral \cite{FP1} (foreseeing herewith no constraint-shell reduction) takes place the current conservation law $\partial_{0}j_{0}^F=\partial_{i}j_{i}^F$, coinciding mathematically with the one in the Gauss-shell reduced {\it equivalent unconstrained system} (EUS), $\partial_{0}j^D_{0}=\partial_{i}j^D_{i}$ for Dirac variables taking on-shell. 
\par
But  the current conservation law $\partial_{0}j^D_{0}=\partial_{i}j^D_{i}$, derived from the {\it classical} equations for the fermionic fields, is destroyed for bound states off-shell, i.e. for "dressed" fermions (and moreover, these bound states are "outside the competence" of the heuristic FP method \cite{FP1}).  
In this context,  the notion "gauge" also concerns the gauge  of sources in  FP path integrals (\ref {fpi11}), but not only the choice of definite Feynman rules (that follows from (\ref{spur2})).
Since  gaugeless (G-invariant) quantization schemes take into account explicitly  the whole physical information from (Gauss law) constraints, it is advantageously to use such G-invariant and relativistic (S) covariant approach to describing composite or collective states. \par 
\medskip
The above sketched quantization scheme by Dirac \cite{Dir} (often referred to as {\it fundamental quantization by Dirac} \cite{Pervush2,LP1,LP2}) is the pattern of such G-invariant and S-covariant quantization schemes.   
As we have made sure above, the Dirac fundamental quantization scheme \cite{Dir, Polubarinov, Pervush2, Nguen2, Heisenberg, Azimov} involves the quantization procedure only for  variables remaining on the constraint-shell reduction of appropriate Hamiltonians and spontaneous violation of initial gauge symmetries (when these  take place). \par \medskip
Now it will be relevant to cite the explicit look of Feynman path integrals \cite{Feynman1, Pervush2,LP1,LP2}, attached to the concrete reference frame (say, the rest reference frame $ l^{(0)}$) and written down in terms of the constraint-shell reduced action functionals (EUS) $W^*$, i.e. in terms of Dirac variables \cite{Pervush2}
\be \label{fi}
 Z_{l^{(0)}}^{*}[ s, {\bar {s}}^*, J^* ]\;=\;\int \prod_{j} DA^D_j
 D\psi^D D{\bar \psi}^D
 e^{iW^{*}[A^D,\psi^D, {\bar \psi}^D] + i S^* },
 \ee
 including the external sources term
\be \label{si}
 S^*\;=\;\int d^4x \left({\bar s}^* \psi^D + {\bar \psi}^D s^*
 +J^*_i A^{Di} \right).
 \ee
The important property of Feynman path integrals is their manifest relativistic covariance \cite{Zumino, Arsen} with respect to the Heisenberg-Pauli relativistic transformations (\ref{dvig}) of  the chosen (rest) reference frame $\eta_0$ 
\cite{Polubarinov, Pervush2, Heisenberg} maintaining the transverse gauge of fields. 
This may be written down as 
\be
\label{Bruno}
Z_{L\eta_0}[s^*,\bar s^*,J^*]= Z_{\eta_0}[L~ s^*,L~\bar s^*,L~J^*].
\ee 
\medskip
To pass then from the Feynman path integral of the form (\ref{fi})    to the FP one, (\ref {fpi11}),  given 
in the fixed gauge $F(A)=0$, one would \cite{Pervush2}:\par
$*)$ replace the variables;\par
$**)$ replace the sources. \par
 The change of variables \rm  is fulfilled by  the Dirac factors $v^T$;
for example,
\be
\label{change1}
A_k^D [A^F]= v^T[A^F](A_k^F+\partial_k)(v^T[A^F])^{-1};
\ee
\be
\label{change2}
\psi^D [A^F]=v^T[A^F]\psi.
\ee 
This change is associated with
the countable number of additional degrees of freedom and  FP determinant ${\rm det}~M_F$ of the transition to new variables of integration. 
These degrees may be removed (to within the Gribov ambiguity in non-Abelian gauge theories \cite{Al.S.,Gribov,Baal, Sobreiro}) by the additional constraint
$F(A)=0$.
Thus the \it constraint-shell \rm functional $Z ^*_{l^{(0)}}$, (\ref{fi}), takes the equivalent form of the FP path integral \cite{Pervush2,LP2}
\begin{eqnarray}\label{fpi}
 Z^{*}[s^*, {\bar s}^*, J^*]\;=\;\int \prod_{\mu}DA^F_\mu D
 \psi^F D{\bar \psi}^F M _{F}
 \delta (F(A^F)) e^{iW[ A^F,\psi^F, {\bar \psi}^F ] + S^*},
 \end{eqnarray} 
where now all  gauge factors $v^T[A^F]$ are concentrated in the source term \cite{Pervush2,Azimov} \footnote{ Indeed \cite{Azimov},
$$ {\bar {s}}^*= {\bar {s}}^F v^T[A^F]; \quad     s^*= (v^T)^{-1}[A^F] ~ s^F; \quad  \bar \psi^D= \bar \psi^F \cdot
 (v^T)^{-1}[A^F].   $$}:
\be\label{fpis}
S^*\;=\;\int d^4x ~ \left( v[A^F] {\bar s}^* \psi^F +
 {\bar \psi}^F  (v[A^F])^{-1}s^*
 +J^*_i A^*_i[A^F])\right).
\ee
Finally, the removal of gauge (Dirac) factors $v^T[A^F]$ by the replacement of gauge fields $A^D \Longrightarrow A$, accompanied by the  change of sources (the step $**)$),
\be
\label{fpso}  
 S^* ~\Rightarrow~S^F=\int d^4x \left( {\bar s}^F \psi^F +  {\bar \psi}^F s^F  + A^F_{\mu}J^{\mu}\right),  \ee
restores the initial FP path integral (\ref{fpi11}) in the considered  gauge theory. 
 Such a  replacement is made
with the only purpose to remove the dependence
of  path integrals on a  reference  frame and initial data. 
But losing the dependence of a gauge model on any  reference  frame is often fraught with serious problems for such a gauge model. 
So, for instance, Schwinger in his paper   \cite{Schwinger} warned  that gauges independent of a reference frame may be physically inadequate to the fundamental operator quantization \cite{Dir}; i.e.  they may distort the spectrum of the original system \footnote{"We reject all Lorentz gauge formulations as unsuited to the role of providing the fundamental operator quantization "  \cite{Schwinger}. }.\par
The situation with asymptotical bound and collective vacuum states, as discussed in Section 2 and involving violating  the {gauge equivalence}  theorem \cite{Taylor, Slavnov1, Arsen}, visually confirms  this warning by Schwinger.
\section{Dirac fundamental quantization of Minkowskian non-Abelian gauge models}
In this section we  give  a short historical retrospective of the development of the  Dirac fundamental quantization method \cite{Dir} in the Minkowskian non-Abelian gauge theory. 
The role of collective vacuum excitations (involving various vacuum rotary effects) in constructing a consequent non-Abelian (Minkowskian) gauge model was considered for the first time in the paper \cite{Pervush1}.\par

The case of such collective vacuum excitations  is just  one of cases (\ref{spur2}) when the {gauge equivalence}  theorem \cite{Taylor, Slavnov1, Arsen} regarding the "heuristic" FP \cite{FP1} and Dirac fundamental \cite{Dir} quantization approaches is violated. 
In the paper \cite{Pervush1}  there was assumed  that in the (Minkowskian) non-Abelian models possessing the strong coupling (YM, QCD), collective vacuum degrees of freedom and long-range correlations of local excitations are possible, similar to those taking  place in the liquid helium  theory \cite{N.N.}. 
Moreover, drawing further a parallel between the (Minkowskian) non-Abelian models and liquid helium  theory \cite{N.N.}, it was concluded about manifest superfluid properties of the physical vacuum in Minkowskian gauge models (indeed \cite{rem1},  such conclusion is correct only for the enough narrow class of Minkowskian gauge models involving vacuum BPS monopole solutions \cite{Al.S.,BPS,Gold} when the initial gauge symmetries are violated and Higgs modes appear as the sign of  such breakdown; we shall discuss this below). 
In \cite{Pervush1} it was demonstrated that the manifest superfluid properties of the  Minkowskian non-Abelian physical vacuum in such models are quite compatible with the Dirac fundamental quantization \cite{Dir}  involving fixing the Coulomb (transverse) gauge for fields. 
As a result, non-Abelian gauge  fields were transformed into (topologically degenerated) Dirac variables satisfying the Coulomb gauge: G-invariant and 
S-covariant simultaneously \cite{Pervush2,Nguen2}.
The Gribov ambiguity \cite{Al.S., Gribov, Baal, Sobreiro} in specifying non-Abelian (transverse) gauge  fields induces in the Minkowskian non-Abelian theory the appropriate second-order differential equation in partial derivatives ({\it the Gribov ambiguity equation} \cite{Pervush2,LP1,LP2}) imposed onto the Higgs field $\Phi$;  this equation proves to be responsible for superfluid properties of the Minkowskian non-Abelian physical (topologically degenerated) vacuum quantized in the Dirac fundamental scheme \cite{Dir}.\par
\medskip
This method of constructing Dirac variables turns the appropriate Gauss law constraint into a homogeneous equation of the form \cite{Pervush1}
\be \label{koop}
(D^2)^{ab} \Phi _b=0
\ee
(with $D$ being the [covariant] derivative),
involving the nontrivial collective vacuum dynamics (more exactly, collective rotations of the Minkowskian non-Abelian vacuum). 
In the paper \cite{Pervush1} it was  postulated the existence of a dynamical variable (denoted as $c(t)$ in \cite{Pervush1})  is responsible for this  collective vacuum dynamics.  
The nature of this variable was explained. 
 The possibility to express $c(t)$ through the integer degree of the map (Pontruagin number) by multiplying it by 
$$
 \int \limits_{n(t_{\rm in})}^ {n(t_{\rm out})} dt $$
was demonstrated (herewith it may be set \cite{Pervush3} $ t_{\rm in, ~out}=\pm T/2$, while interpreting $c(t)$ as a noninteger  degree of map becomes transparent).
The necessity to take account of theoretical-group properties of the considered  Minkowskian non-Abelian model is the basis for such form                 of the dynamical cooperative variable $c(t)$ (as well as of another dynamical variables that this model implicates).
\par 
This allowed to write down explicitly the item in the YM Lagrangian  describing the collective vacuum rotations \cite{Pervush1}
\be \label{rotary}
L_{\rm coop}= [\int d^3 x (D_i \Phi)^2] \frac 1 2 \dot c^2(t).
\ee 
 The similar nature of these  collective vacuum rotations in Minkowskian non-Abelian models and quantum vortices in a liquid helium specimen was noted (see e.g. \S\S 30-31 in \cite{Halatnikov}).  
It was shown in \cite{Pervush1} that the collective vacuum rotations (involving the appropriate rotary item $ L_{\rm coop}$ (\ref{rotary}) in the YM Lagrangian)  may be expressed in terms of Higgs vacuum modes $\Phi^a$, setting the transverse vacuum "electric" field $D_\mu E^\mu =0$. The connection between the zero modes $Z^a\sim \dot c(t) \Phi^a $ of the YM Gauss law constraint  and this transverse vacuum "electric" field $E$ was ascertained.\par 
Additionally, there was demonstrated that the {\it purely real} and simultaneously discrete energy-momentum spectrum 
\be \label{rspec} P\sim 2\pi k+ \theta; \quad k\in {\bf Z};\ee
corresponds to the collective vacuum rotations in the Minkowskian non-Abelian theory.  
This purely real and discrete energy-momentum spectrum is the alternative to the 
{\it complex} topological momentum
\be 
\label{tm}
P_{\cal N}=2\pi k \pm 8\pi i/g^2\equiv 2\pi k +\theta \ee
proper (as it was demonstrated in the papers \cite{Pervush1,Galperin} and then repeated in Ref. \cite{Arsen}; see also \cite{rem3})  to the Euclidian $\theta$-vacuum.  
This  result \cite{Arsen, Pervush1,Galperin} means, as it is easy to see, that topologically degenerated instanton solutions inherent in 
the Euclidian YM model \cite{Al.S.,Cheng, Bel} are purely gauge, i.e. {\it unphysical and unobservable}, fields. 
The additional argument in favour of the latter assertion was made recently in Ref. \cite{LP1}.
It was noted that the $\theta$-vacuum
plane wave function \rm \cite {Pervush1}
\be
\label{plan}
\Psi _0[A]=\exp (iP_{\cal N}X[A]),
\ee
corresponding to the zero energy $\epsilon=0$ of an instanton \cite{Cheng, Bel} (with $ X[A]$ being the winding number functional taking integers), is specified wrongly at the minus sign before $P_{\cal N}$ in  (\ref{plan}). 
This implies that it is impossible to give the correct probability description of the instanton $\theta$-vacuum \cite{Al.S.,Cheng, Bel} \footnote{since the Hilbert space of (topologically degenerated) instanton states becomes non-separable in this case.}; that is why the latter one refers  to unobservable, i.e. {unphysical}, values.

In Ref. \cite{Arsen} (see also \cite{rem3}) the result (\ref {tm})  was referred to as the so-called \it no-go theorem\rm: the presence of unphysical solutions in the Euclidian 
instanton YM (non-Abelian) theory \cite{Bel}.

\medskip
Later on, in Ref. \cite{Pervush3}, it was explained the common property of cyclical motions (to which belong also the collective vacuum rotations inside the Minkowskian non-Abelian vacuum described in Ref. \cite{Pervush1}):  that all they possess  the discrete energy-momentum spectrum, similar to described above. 
This can serve as a definition of the Minkowskian $\theta$-vacuum, somewhat alternative to that of Ref. \cite{Callan1} given for the $\theta$-vacuum in the Euclidian non-Abelian theory \cite{Bel} involving instantons (the arguments \cite{Callan1} were then repeated in Refs. \cite{Arsen, Pervush1}). 
The discrete energy-momentum spectra $P$ of cyclical motions found \cite{Pervush3} to be, firstly, a purely quantum effect disappearing in the 
semi-classical limit $\hbar \to 0$ and, secondly, such motions cannot vanish until $\theta\neq 0$.\par
Really, in the $\hbar$ terms, the discrete energy-momentum spectra $P$ of cyclical motions may be expressed as \cite{Pervush3}
$$ P=   \hbar ~\frac {2\pi k+ \theta} L, $$
with $L$ being the length of the whole closed line along which a physical material point (say, physical field) moves. 
Thus when $\theta\neq 0$, the momentum $P$ attains its nonzero minimum
$P_{\rm min} = \hbar \theta /L$ as  $k=0$.\par
This is the display of the so-called {\it Josephson effect} \cite{Josephson} for superconductors including in an electric circuit. 
The essence of this effect just in the persistent cyclical motion of a quantum "train", cannot stop until $\theta\neq 0$ \cite{Pervush3}.\par
For the Minkowskian non-Abelian physical vacuum such Josephson effect comes to the vacuum (transverse) 
 "electric" field $E$, proving to be a definite function of the appropriate rotary 
energy-momentum spectrum $P$. More exactly, $E=f(k, \hbar, \theta)$.  
This means that $E$ also attains its nonzero minimum value $E_{\rm min}$ as $k=0$ and $\theta\neq 0$. \par
The dependence of $E_{\rm min}$ on the Planck constant $\hbar$ (this was noted for the first time in Ref.  \cite{Pervush3}) is connected with the claim for the strong interaction coupling constant to be, indeed, dimensionless; that results $g^2/(\hbar c)^2$ in the lowest order of the perturbation theory. 
In this case \cite{Pervush3}, the collective rotations term in the 
non-Abelian action functional proves to be directly proportional to the Planck constant $\hbar$ and disappearing in the \linebreak (semi)classical limit $\hbar\to 0$ (see below). \par
\bigskip
As we have already discussed in the previous Section, the general principles for constructing constraint-shell (Gauss-shell) gauge models were stated in  Refs. \cite{Nguen2,Azimov}.
These general principles (with some correctives: for instance, 
replacing $\partial$ by the [covariant] derivative $D$) may be spread from 
four-dimensional constraint-shell QCD to the \linebreak (Minkowskian) non-Abelian gauge models (including that involving Higgs and fermionic modes and violating initial symmetries groups). 
A remarkable feature of the constraint-shell reduction of gauge models proves to be the appearance of current-current instantaneous interaction therms in  EUS Hamiltonians. 
For comparison, in the four-dimensional constraint-shell QCD, the current-current instantaneous interaction term in the appropriate Gauss-shell reduced Lagrangian density ${\cal L}^D(x)$ is read as \cite{Nguen2}
\be \label{QEDt1}  \frac{1}{2} j_0^D \frac{1}{\Delta} j_0^D    \ee
and implicates G-invariant currents
$$ j_\mu^D=  e\bar \psi^D \gamma_\mu \psi^D.  $$ 
In  the (Minkowskian) non-Abelian constraint-shell QCD the analogy of (\ref{QEDt1})  the "potential" term \cite{Werner,David1} will be
\be
\label{cint}
\frac {1}{2} \int \sb{V_0} d^3 x d^3 y j_{{\rm tot},(0)} ^b ({\bf x}) G_{bc} ({\bf x},{\bf y}) j_{{\rm tot},(0)}^c({\bf y}).
\ee
in the constraint-shell reduced QCD Hamiltonian.

This "potential" term involves \cite{Werner,David1}
the topologically trivial and G-invariant  total currents
\be
\label{tocar}
j_{{\rm tot},(0)}^a= g\bar \psi^I (\lambda ^a /2)\gamma_0  \psi^I +\epsilon ^{abc} \tilde E_i^{T b} \tilde A^i_{T c(0)},
\ee
involving fermionic topologically trivial  Dirac variables $\psi^I$, $\bar \psi^I$.

In this equation, the transverse "electric" tension $ D^i \tilde E_i^{T a}=0$, belonging to the excitation spectrum of (Minkowskian) constraint-shell QCD, can be expressed \cite{Pervush1} through the topologically trivial gauge potentials $ \tilde A^i_{ a(0)}$ (which can be chosen to be also transverse: $ D_i \tilde A^i_{T a(0)}=0$; for instance, the Dirac variables  (\ref{Dv}))
\be \label{trep}
\tilde E_i^{(T)a} =(\delta_{ij} -D_i \frac{1}{D^2})^{ab} \partial_0 \tilde A_{ib}. 
\ee 
The Green function $ G_{ab} ({\bf x},{\bf y})$ of the Gauss law constraint \cite{Pervush2,LP1,LP2}
\be
\label{cur-t}
D_i^{cd}(A)D^i_{db}(\Phi^{(0)})\tilde \sigma ^b = j_{{\rm tot}(0)} ^c
\ee
enters the "potential term" (\ref{cint}).

The longitudinal  "electric" field $\tilde \sigma ^a$ has the form \cite{Pervush2}
\be
\label{sigma}
 \sigma ^a[A^T,E^T]= (\frac {1}{D_i (A)\partial^i})^{ac}\epsilon_{cbd}A_k^{Tb}E^{Tkd}
\ee 
and involves transverse fields $A^T$ and $E^T$.
This equation reflects the manifest nonlinear nature of non-Abelian gauge models of the YM type.

\medskip
Speaking about Minkowskian constraint-shell QCD, one should note the  support of the infrared quark confinement in that model.
It turned out that the infrared quark confinement in Minkowskian constraint-shell QCD has actually topological origins. 

   In Refs. \cite{Nguen2,Azimov}   the interference of topological Gribov multipliers \cite{Gribov} in the gluonic and fermionic Green functions in all  orders of the perturbation theory was demonstrated. 
More exactly,  these  stationary gauge multipliers of the typical form $v^{T(n)}({\bf x})$, depending explicitly on topologies $n\in {\bf Z}$,  enter {\it topological Dirac variables} in non-Abelian gauge models
\be \label{staz} v^{T(n)}({\bf x})=  v^{T(n)}({\bf x},t)\vert_ {t= t_0}.  \ee
In non-Abelian gauge theories matrices $ v^{T(n)}({\bf x},t)$ may be found easily, satisfying the  Cauchy condition (\ref{staz}) and Eq. (\ref {udalenie}) \cite{David3} specifying the (topological) Dirac variables (\ref {per.Diraca}) in these theories quantized by Dirac \cite{Dir}. 

As it was shown in Ref. \cite{LP1,Pervush1},
\be \label{dress}
v^ {T(n)}(t,{\bf x})= v^{T(n)} ({\bf x})T \exp \{\int  \limits_{t_0}^t 
[\frac {1}{D^2} \partial_0 D_k {\hat A}^k]~d\bar t ~\}, \ee
where  the symbol $T$  stands for the  time ordering of the matrices under the exponent sign. 

Following Ref. \cite{LP1}, one 
 notes the exponential expression in (\ref{dress}) as $U^D[A]$; this expression may be rewritten \cite{LP1} as
\be
\label{UD}
U^D[A]= \exp \{\frac {1}{D^2} D_k  {\hat A}^k\}.
\ee
\medskip
This mechanism \cite{Azimov} of  the infrared (at the spatial infinity $\vert {\bf x}\vert \to \infty$) ({\it destructive}) interference of Gribov stationary multipliers $ v^{T(n)}({\bf x},t)$ in the gluonic and fermionic Green functions in all the orders of the perturbation theory leads to the
following results.

Firstly, one claims \cite{Arsen,Azimov,Pervush3}
\be \label{bondari} v^{T(n)}({\bf x})\to \pm 1, \quad {\rm as} ~~\vert {\bf x}\vert \to \infty.\ee
This claim imposed onto the Gribov stationary multipliers $ v^{T(n)}({\bf x},t)$ at the spatial infinity is quite natural and legitimate.

So in the Euclidian instanton model \cite{Bel} the similar spatial asymptotic of gauge matrices is equivalent to disappearing instantons  at the "four-dimensional" infinity $\vert x\vert \to\infty$ (as it was noted, for instance, by Dashen et al \cite{Bel})
\be \label{asi}  A_\mu  (x)\to 0, \quad \vert x\vert \to\infty. \ee 
As a consequence, the Pontruagin degree of the map \cite{Al.S.}, 
$$  n(g)=\frac {1}{24\pi^2} \int \sb {S^3} ~ {\rm tr}~ \{ (g^{-1}(x)dg(x))  
[ g^{-1}(x)dg(x) \wedge g^{-1}(x)dg(x)]\}dx, 
      $$
involving gauge matrices $g(x)$, takes integer values.

Indeed, disappearing  gauge fields (\ref{asi}) at the "four-dimensional" infinity has the universal nature for the Euclidian as well as  for the Minkowskian space-time: in particular, the boundary condition (\ref{bondari}) \cite{Arsen,Azimov,Pervush3}  imposed onto the Gribov stationary multipliers $ v^{T(n)}({\bf x},t)$ at the spatial infinity in the Minkowskian gauge model (quantized by Dirac \cite{Dir}) is quite correct.

\medskip
Secondly, as  was shown in  Ref. \cite{Azimov}, in the lowest order of the perturbation theory, averaging (quark) Green functions over all  topologically nontrivial field configurations (including vacuum monopole ones, us discussed below) results in \cite{Nguen2,Azimov} 
\be \label{oneqw}
G({\bf x},{\bf y})= \frac {\delta}{\delta s^*(x)} \frac {\delta}{\delta \bar
s^*(y)} Z_{\rm conf} (s^*, \bar s^*,J^*)\vert_{ s^*= \bar s^*=0} =G_0(x-y) f({\bf x},{\bf y}),
\ee
with $ G_0(x-y)$ being  the (one-particle) quark propagator in the perturbation theory and
\be \label{interf}
f({\bf x},{\bf y})= \lim_{\vert {\bf x}\vert \to \infty, ~ \vert {\bf y}\vert \to \infty} \lim_{L \to \infty} (1/L) \sum \limits _{n=-L/2}^{n=L/2} v^{(n)} ({\bf x}) v^{(n)} ({\bf -y}). \ee 
The origin of the generating functional $Z_{\rm conf} (s^*, \bar s^*,J^*)$, entering (\ref{oneqw}), is following.
It comes from the standard FP integral (\ref {fpi11}) \cite{FP1} in which one fix the transverse gauge for (YM) fields $A$ \cite{Pervush2, Pervush1}:
\be \label{tg} D_i (A) \partial_0A^i =0,  \ee
turning these fields into (topological) Dirac variables of the (\ref {per.Diraca}) type.

In this case \cite{Gitman} the FP operator $M_F$ takes the look $\Delta _{\rm FP}$ \cite{Baal, Sobreiro}, (\ref {FP operator}):
\be 
\label{FP1}
\hat \Delta = -(\partial_i D^i (A)) = -( \partial_i^2 + \partial_i~ {\rm ad} (A^i))
\ee 
with
$$ {\rm ad} (A) X \equiv [A,X]$$ 
for an element $X$ of the appropriate Lee algebra.  
Such form of the FP operator is mathematically equivalent to (\ref {FP operator}) \cite{Baal, Sobreiro} at setting $A_0=0$ for temporal components of gauge fields.
In particular, it becomes correctly for the removal (\ref {udalenie}) \cite{Dir,David3} of these components in the  Dirac fundamental quantization scheme.

\medskip
Upon fixing the gauge (\ref {tg}), the FP integral (\ref {fpi11}) takes the form \cite{Azimov}
$$
Z_{R,T} (s^*, \bar s^*,J^*)= \int D A_i^* D\psi^* d \bar\psi^* ~{\rm det}~ \hat \Delta ~\delta (\int \limits_ {t_0}^t d\bar t D_i (A)\partial_0 A^i)
$$
\be \label{Zr}
\times
\exp \{ i \int \limits_ {-T/2}^ {T/2} dt \int \limits_ {\vert {\bf x}\vert \leq R} d^3x [{\cal L} ^I (A^*, \psi^*) +  \bar s^*\psi^* +\bar \psi^* s^* +J^{*a}_i A^{i*}_a]\}. 
\ee 
It includes the Lagrangian density ${\cal L}^I$ \cite{Nguen2} corresponding to the constraint-shall action of the Minkowskian non-Abelian theory (Minkowskian QCD) taking on the surface of the Gauss law constraint (\ref {Gau}); here $R$ is a large real number, and one can assume  that $R\to \infty$. \par

Thus in the case (\ref{Zr}) of the constraint-shall Minkowskian non-Abelian theory, when the transverse gauge (\ref {tg}) is fixed, turning gauge fields into (topological) Dirac variables $A^*$, this FP integral depends formally on these Dirac variables and also on $\psi^*$, $\bar\psi^*$. 
Then the generating functional $Z_{\rm conf} (s^*, \bar s^*,J^*)$, entering Eq. (\ref {oneqw}) \cite{Azimov} specifying quark Green functions in Minkowskian QCD involving topologically nontrivial (vacuum) configurations, may be derived from the FP integral (\ref{Zr}) by its averaging over the Gribov topological degeneration \cite{Gribov} of initial data, i.e. over the set $\bf Z$ of integers
\be \label{Zcon}
Z_{\rm conf} (s^*, \bar s^*,J^*)= \lim _{\vert {\bf x}\vert\to \infty,~T\to \infty} \lim _{L\to \infty} \frac{1}{L} \sum \limits _{n=-L/2}^{n=L/2}  Z_{R,T}^I (s^*_{n,\phi_i}, \bar s^*_{n,\phi_i},J^*_{n,\phi_i}), \ee 
with $ Z_{R,T}^I (s^*_{n,\phi_i}, \bar s^*_{n,\phi_i},J^*_{n,\phi_i})$ being the FP path integral (\ref{Zr}) rewritten in terms of Gribov exponential multipliers $v^{T(0)}({\bf x})$ \footnote{It is correct due to the manifest G-invariance of the constraint-shell theory.
For the same reason, the constraint-shell Lagrangian density ${\cal L}^I$, entering the FP integral (\ref{Zr}), may be expressed solely in terms of  topologically trivial Gribov exponential multipliers $v^{T(0)}({\bf x})$ \cite{Azimov}. }.

The variation of $ Z_{\rm conf} (s^*, \bar s^*,J^*)$ by the sources,
$$ (\prod \limits _{\alpha=1}^3 \frac{\delta}{\delta s^*_{n,\phi_\alpha}})(\prod \limits _{\beta=1}^3 \frac{\delta}{\delta \bar s^*_{n,\phi_\beta }})(\prod \limits _{\gamma=1}^3 \frac{\delta}{\delta J^*_{n,\phi_\gamma }}),$$
involving the appropriate Euler angles $\phi_\alpha (x_\alpha)$ ($\alpha =1,2,3$; $ x_\alpha $, $ y_\alpha$, $ z_\alpha $ are the Cartesian coordinates) and topological charges $n\in \bf Z$, just results the Green functions of the (\ref {oneqw}) \cite{Azimov} type (in particular, to derive Eq. (\ref {oneqw}) for fermionic Green functions, it is necessary to omit the variation of $ Z_{\rm conf} (s^*, \bar s^*,J^*)$ by gauge currents $ J^*_{n,\phi_\alpha }$).

\medskip
Returning to Eq. (\ref {interf}), note that always $ f({\bf x},{\bf y})=1$ due to the spatial asymptotic (\ref {bondari}) for the Gribov topological multipliers $v^{T(n)}({\bf x})$.
This implies that only "small" (topologically trivial) Gribov exponential multipliers  $v^{T(0)} ({\bf x})$ contribute in $ f({\bf x},{\bf y})=1$ and, therefore, in the (one-particle) quark Green function (\ref{oneqw}).
A similar reasoning \cite{Azimov} remains correct also for multi-particle quark and gluonic Green functions in all  orders of the perturbation theory. 

Just  above described   only "small" surviving Gribov exponential multipliers  $v^{T(0)} ({\bf x})$ in Green functions in all  orders of the perturbation theory was referred to as the (infrared) {\it topological confinement} in the series of papers (for instance, in the review \cite{Pervush2}, multiply cited  in the present study). 

\bigskip
The new stage in the development of the Minkowskian non-Abelian  model quantized in the fundamental scheme by Dirac \cite{Dir} began in the second half of the 90-ies and continues  to date.
It is connected with the papers \cite{Pervush2,LP1, David3, LP2, David1}. 
These papers were devoted to the Dirac fundamental quantization \cite{Dir} of the Minkowskian non-Abelian  theory involving the spontaneous breakdown (say, $SU(2)\to U(1)$) of the initial gauge symmetry and appearing Higgs modes (we shall refer to such theory as to the {\it Minkowskian Higgs non-Abelian model}).

Now we should like dwell on the some points of the recent investigations \cite{Pervush2,LP1, David3, LP2, David1} (especially on those  representing new results in comparison with the "old" research about the Dirac fundamental quantization of the Minkowskian non-Abelian  model).

\bigskip

A. {\it Vacuum BPS monopoles}.

The idea to utilize vacuum BPS monopole solutions \cite{Al.S., BPS, Gold} for describing \linebreak Minkowskian Higgs models quantized in the fundamental scheme by Dirac \cite{Dir} was proposed, probably, already in the paper \cite{Arsen}.
In the recent papers \cite{Pervush2,LP1, LP2, David1} this idea becomes the basic one, while in the work  \cite{David3} the spatial asymptotic of vacuum BPS monopole solutions in the shape of Wu-Yang monopoles \cite{Wu} was studied \footnote{ Wu-Yang monopoles \cite{Wu} are solutions to the classical equation of motion 
$$ D^{ab}_k(\Phi_i)F^{bk}_a(\Phi_i)=0       $$
of the  "pure" (Minkowskian) YM theory (without Higgs fields).
The solution to this classical equation is the "magnetic" tension $ F^{bk}_a $ taking the form 
$$   B^{i a}(\Phi_i)= \frac {x^ax^i}{gr^4}.$$   

Thus such "magnetic" tension diverges at the origin of coordinates $r\to 0$, while the spatial YM components
$$  {\hat\Phi}_i =-i \frac {\tau ^a}{2}\epsilon _{iak}\frac {x^k}{r^2}f^{\rm WY}(r),    $$
with $ f^{\rm WY}(r)=\pm 1$, correspond to Wu-Yang monopoles \cite{Wu} with topological charges $\pm 1$, respectively.

Indeed, the above classical equation of motion implies  the following equation for the function $f(r)$ \cite{Pervush2,LP1}:
$$  D^{ab}_k(\Phi_i)F^{bk}_a(\Phi_i)=0 \Longrightarrow \frac {d^2f}{d r^2}+\frac {f (f^2-1)}{r^2}=0.
      $$
Herewith $f(r)=\pm 1$ just results Wu-Yang monopoles \cite{Wu} with topological charges $\pm 1$: $f(r)= f^{\rm WY}(r)$, while the solution $ f(r)\equiv f^{\rm PT}= 0$ corresponds  to the naive unstable perturbation theory, involving  the asymptotic freedom formula \cite{Gr}.}.

Unlike the Wu-Yang monopoles \cite{Wu} (us analysed above briefly), diverged as $1/r$ at the origin of coordinates, YM vacuum BPS monopole solutions \cite{Al.S., BPS, Gold} \linebreak $ A^a_{i} (t,{\bf x})\equiv\Phi^{aBPS}_i({\bf x}) $ (in denotations \cite{LP1}) are regular in the whole spatial volume.
Thus a good approximation of Wu-Yang monopoles \cite{Wu} by YM vacuum BPS monopoles \cite{Al.S., BPS, Gold} is on hand.

 By the way, note  that Euclidian YM instanton solutions $A_i^a$ \cite{Bel} are also singular  at the origin of coordinates (for instance, this was demonstrated in the monograph \cite{Al.S.}, in \S$\Phi$23).

As to the Higgs vacuum BPS monopole solutions $\Phi^a ({\bf x})    $ \cite{LP1}, they diverge at the spatial infinity although they are regular at the origin of coordinates (like YM BPS monopoles $\Phi^{aBPS}_i({\bf x})$) \footnote
{Meanwhile, as  shown in Ref. \cite{BPS}, the vacuum "magnetic" field $\bf B$ corresponding  to the vacuum YM BPS monopole solutions $\Phi^{aBPS}_i({\bf x})$ diverges as $1/r^2$ at the origin of coordinates, and in this, as the author of the present work recognizes, is a definite problem  requiring a solution.
Really, in this ultraviolet region of the momentum space,  gluons and quarks would be asymptotically free \cite{Cheng, Gr}, but the $1/r^2$ behaviour of the vacuum "magnetic" field $\bf B$ hinders  this.}.

\medskip
The important new step in research of vacuum  BPS monopole solutions undertaken in  resent papers \cite{LP1,LP2}, in comparison with "classical" issues \cite{Al.S., BPS, Gold}, is assuming  (topologically degenerated) BPS monopole solutions (in the YM and Higgs sectors of the Minkowskian non-Abelian Higgs model) depending explicitly on the effective Higgs mass $m/\sqrt \lambda$ taken in the {\it Bogomol'nyi-Prasad-Sommerfeld} (BPS)  limit  \cite{Al.S., LP1,LP2,BPS, Gold} 
$$ m\to 0; \quad  \lambda \to 0 $$
for the Higgs mass $m$ and  Higgs selfinteraction $\lambda$, respectively.

More exactly,  this dependence of  vacuum  BPS monopole solutions on the effective Higgs mass $m/\sqrt \lambda$ is reduced to their dependence on the value $\epsilon$ introduced as \cite{LP1,LP2}
\be 
\label{lim} 
\frac{1}{\epsilon}\equiv\frac{gm}{\sqrt{\lambda}}\not =0.
\ee
Since the YM coupling constant $g$ and Higgs selfinteraction constant $\lambda$ are  dimensionless,  $\epsilon$ has the dimension of distance.
It may be treated as the effective size of  vacuum  BPS monopoles and proves to be inversely proportional to the spatial volume $V$ occupied by the (YM- Higgs) field configuration, as  was shown in Ref. \cite{LP1,LP2}
\be 
\label{masa}  \frac{1}{\epsilon}=\frac{gm}{\sqrt{\lambda}}\sim \frac{g^2<B^2>V}{4\pi}, \ee 
with $<B^2>$ being the vacuum expectation value of the "magnetic" field $\bf B$ set by the Bogomol'nyi equation \cite{Al.S., LP1,LP2,BPS, Gold}
\be
\label{Bog} 
{\bf B} =\pm D \Phi.
\ee 
Thus one can speak that the effective size $\epsilon$ of  vacuum  BPS monopoles is a function of the distance $r$ with the inversely proportional dependence 
$$ \epsilon (r)\equiv f(r)\sim O(r^{-3}).  $$ 
\medskip
There is an important physical meaning of the  effective size $\epsilon$ of  vacuum  BPS monopoles and the effective Higgs mass $m/\sqrt \lambda$. 
As  follows from (\ref{masa}), the  values are  some functions of the distance $r$, and  this gives the possibility to utilize them as scale parameters describes renormalization group (RG) properties  of the Minkowskian non-Abelian Higgs model (quantized  by Dirac \cite{Dir}).
For instance, the  effective Higgs mass $m/\sqrt \lambda$ may be treated as a {\it Wegner mass} \cite{Kadanoff, Wegner}.
The possibility to apply  the  Bogomol'nyi equation (\ref{Bog}) for the Dirac fundamental quantization \cite{Dir} of the Minkowskian Higgs model was pointed out already in the paper \cite{Arsen}.

In  recent articles  \cite{LP1,LP2,rem1}, the relation of the Bogomol'nyi equation (\ref{Bog}) and vacuum BPS monopole solutions with superfluid properties of the Minkowskian Higgs model quantized by Dirac was noted (as we have pointed out above, such superfluid properties of that model were assumed already in the paper \cite{Pervush1} \footnote{This work summarized the series of results \cite {Pervushin}.}).
The physical  non-Abelian vacuum specified by YM and Higgs vacuum BPS monopoles is described by the Bogomol'nyi equation (\ref{Bog}) as a potential superfluid liquid similar to the superfluid component in a liquid helium II specimen \cite{N.N.} \footnote{More precisely, one can trace easily a transparent parallel between the vacuum "magnetic" field $\bf B$ set by the Bogomol'nyi equation (\ref{Bog}) and the velocity ${\bf v}_s$ \cite{Landau52} of the superfluid motion in a liquid helium II specimen
$$  {\bf  v}_s =\frac{\hbar}{m} \nabla \Phi(t,{\bf r}),         $$
with $m$ being the mass of a helium atom and $\Phi(t,{\bf r})$ being the phase of the helium Bose condensate wave function $\Xi (t,{\bf r})$
$$  \Xi (t,{\bf r})= \sqrt {n_0(t,{\bf r})}~ e^{i\Phi(t,{\bf r})},     $$
where $ n_0(t,{\bf r})$ is the number of particles in this helium Bose condensate.}.

\medskip
On the other  hand, although manifest superfluid properties of the Minkowskian Higgs model are proper only at utilizing BPS monopole solutions \cite{Al.S., LP1,LP2,BPS, Gold}
for describing the appropriate (physical) vacuum, the Bogomol'nyi equation (\ref{Bog}) is associated, indeed, with the FP "heuristic" quantization \cite {FP1} of that model.
As it was demonstrated for instance in Ref.  \cite{Al.S.} (in \S$\Phi$11), the Bogomol'nyi equation (\ref{Bog}) can be derived without solving the YM Gauss law constraint (\ref{Gau}), but only  evaluating the {\it Bogomol'nyi bound} \cite{Al.S.,LP1,LP2}
\be 
\label{Emin}
E_{\rm min}= 4\pi {\bf m }\frac {a}{g},~~~~~~~~~~~~\,\, ~~~~~a\equiv \frac{m}{\sqrt{\lambda}}
\ee  
(where $\bf m$ is the magnetic charge) of the (YM-Higgs) field configuration involving vacuum BPS monopole solutions taken in the BPS limit.

By applying the Dirac fundamental quantization scheme \cite{Dir} to the Minkowskian non-Abelian Higgs model implicating BPS monopole solutions, the potentiality and superfluidity proper to the physical  vacuum of that model are set \cite{rem2}  by the Gribov ambiguity equation, coinciding mathematically with (\ref {koop}) (we have already discussed this at the beginning of the present section).
In this case the connection between the Bogomol'nyi equation (\ref{Bog}) and the Gribov ambiguity equation (having the form (\ref {koop})) is accomplished via the Bianchi identity  $ D~ B=0$.

In the papers \cite{Pervush2,LP1,LP2,David1} the solution to the Gribov ambiguity equation was found in the form of the so-called {\it Gribov phase}
\be
 \label{phasis}
{\hat \Phi}_0(r)= -i\pi \frac {\tau ^a x_a}{r}f_{01}^{BPS}(r), \quad f_{01}^{BPS}(r)=[\frac{1}{\tanh (r/\epsilon)}-\frac{\epsilon}{r} ].
 \ee 
It is the $U(1)\to SU(2)$ isoscalar "made" of vacuum Higgs BPS monopole solutions.

This allowed to write down explicitly Gribov topological multipliers $ v^{T(n)}({\bf x})$, (\ref{staz}), through the Gribov phase (\ref {phasis}) 
\be
 \label{mon.deg}
v^{T(n)}({\bf x})=\exp [n\hat \Phi _0({\bf x})].
\ee
Thus the immediate connection between the Dirac fundamental quantization \cite{Dir} of the Minkowskian non-Abelian Higgs model involving vacuum BPS monopole solutions \cite{LP1,LP2} (this quantization comes to topological Dirac variables (\ref {per.Diraca}), G-invariant and taking in the transverse gauge (\ref {Dv}), as solutions to the YM Gauss law constraint (\ref{Gau})), the Gribov ambiguity equation and the Bogomol'nyi equation (\ref{Bog}) (responsible for manifest superfluid properties of that model) was ascertained.

Additionally, the function $ f_{01}^{BPS}(r)$ entering the expression for the Gribov phase (\ref {phasis}) has the spatial asymptotic 
$$  f_{01}^{BPS}(r) \to 1 \quad  {\rm as}~ r\to \infty,  $$
as  shown in the paper \cite{David1}. 

Such spatial asymptotic \cite{David1} of $ f_{01}^{BPS}(r)$  in a good agreement with the boundary condition (\ref {bondari}), should be imposed onto Gribov topological multipliers $ v^{T(n)}({\bf x})$ at the spatial infinity in order to ensure the infrared  topological confinement \cite{Azimov} of   topologically nontrivial  multipliers $ v^{T(n)}({\bf x})$  with $n\neq 0$ in fermionic and gluonic Green functions in all the orders of the perturbation theory. 

\bigskip
B. {\it  specific character of the Josephson effect in the Minkowskian non-Abelian Higgs model.}

In  recent papers \cite{Pervush2,LP1, David3, LP2, rem3, David1} the following specific features of the Josephson effect proceeding in the Minkowskian non-Abelian Higgs model quantized by Dirac \cite{Dir} were noted.
As  demonstrated in Ref. \cite{David3}, the main manifestation of the Josephson effect \cite{Pervush3, Josephson} in the Minkowskian non-Abelian Higgs model quantized by Dirac is the minimum vacuum "electric" field $\bf E$ never vanishing if $\theta \neq 0$: 
\be\label{sem}
(E_i^a)_{\rm min}= \theta \frac {\alpha_s}{4\pi^2\epsilon} B_i^a; \quad -\pi\leq \theta \leq \pi
\ee 
(with $\alpha_s\equiv g^2/4\pi$).

Such minimum value of the vacuum "electric" field $\bf E$ corresponds to trivial topologies $k=0$, while generally \cite{David3},
\be\label{se} F^a_{i0}\equiv E_i^a=\dot c(t) ~(D_i (\Phi_k^{(0)})~ \Phi_{(0)})^a= P_c \frac {\alpha_s}{4\pi^2\epsilon} B_i^a (\Phi _{(0)})= (2\pi k +\theta) \frac {\alpha_s}{4\pi^2\epsilon} B_i^a(\Phi_{(0)}).
\ee
This equation for the vacuum "electric" field $\bf E$ contains (topologically trivial) Higgs vacuum BPS monopoles $\Phi_{(0)}^a$, whereas  the (covariant) derivative $ D_i^a (\Phi_k^{(0)})$ is taken in the background of (topologically trivial) YM BPS  monopoles $\Phi_k^{a(0)}$.

In the papers \cite{LP1, LP2} vacuum "electric" fields $\bf E$ were referred to as vacuum "electric" monopoles.
Their actual form
\be
\label{el.m}
F^a_{i0}\equiv E_i^a=\dot c(t) D ^{ac}_i(\Phi_k ^{(0)})\Phi_{(0)c}({\bf x}),
\quad
 E_i^a\sim  D ^{ac}_i Z_c     \ee 
was elucidated already in the work \cite{Pervush1}.

Herewith Eq. (\ref{se}) \cite{David3} for vacuum "electric" monopoles (\ref {el.m})
follows immediately from the rotary Lagrangian (\ref {rotary}) \cite{Pervush2,LP1, Arsen, David3, LP2, Pervush1, David1} recast into the action functional \cite{David3}
\be
\label{rot} 
W_{\rm coop}=\int d^4x \frac {1}{2}(F_{0i}^c)^2 =\int dt\frac {{\dot c}^2(t) I}{2},
\ee
with 
\be
\label{I1}
I=\int \sb {V} d^3x (D_i^{ac}(\Phi_k)\Phi_{(0)c})^2 = \frac {4\pi^2\epsilon}{\alpha _s} =\frac {4\pi^2}{\alpha _s}\frac {1}{ V<B^2>} \ee
being the {\it rotary momentum} of the Minkowskian (YM-Higgs) vacuum. \medskip 

Since actually \cite{Pervush3,rem3} 
$$ \alpha_s= \frac {g^2}{4\pi (\hbar c)^2},$$
now it becomes obvious that the rotary momentum $I$ and, therefore, the action functional (\ref{rot}), prove to be directly proportional to the Planck constant squared $\hbar^2$. Thus in the (semi)classical limit $\hbar\to 0$ collective rotations of the discussed physical BPS monopole vacuum disappear, as it was already noted.
\medskip 

One obtains directly from (\ref{rot})  that
\be \label{Pcr}  P_c \equiv \frac{\partial W_{\rm coop}}{\partial\dot c} ={\dot c} I= 2\pi k +\theta.         \ee
This confirms the general Eq. (\ref {rspec}) assumed in \cite{Pervush1} for the Minkowskian non-Abelian Higgs theory quantized by Dirac \cite{Dir} (now in the concrete case of vacuum BPS monopole solutions).
In other words, vacuum BPS monopole solutions involve the purely real energy-momentum spectrum of collective rotations associated with the topological dynamical variable $c(t)$.

\medskip
The important point of Eq. (\ref {I1}) \cite{David3} for the rotary momentum $I$ of the Minkowskian (YM-Higgs) vacuum is its direct proportionality to the effective BPS monopole size $\epsilon$, (\ref {masa}).
Thus the contribution of the collective (YM-Higgs) vacuum rotations in the total action of the Minkowskian (Gauss-shell) non-Abelian Higgs theory quantized by Dirac is suppressed in the infinite spatial volume limit $V\to \infty$.
On the other hand, the presence in (\ref {I1}) of the vacuum expectation value $<B^2>$ for the "magnetic" field $\bf B$ is the direct trace therein of vacuum BPS monopole solutions associated with the Bogomol'nyi equation (\ref {Bog}). 

\medskip
As  noted in \cite{David3}, the minimum never vanishing (until $\theta\neq 0$) vacuum "electric" field ${\bf E}_{\rm min}$ (\ref {sem}) and the (constraint-shell) action functional $W_{\rm coop}$ (\ref {rot}), describing the collective rotations of the physical Minkowskian (YM-Higgs) vacuum, are a specific display of the general Josephson effect \cite{Pervush3} in the Minkowskian non-Abelian Higgs model quantized by Dirac.
This effect now comes to the "persistent field motion" around the "cylinder" of the effective diameter $\sim \epsilon$, (\ref {masa}).
Moreover, repeating the arguments \cite{Pervush3} regarding the Josephson effect, there was shown in \cite{David3} that
\be 
\label{per}
\Psi_{c} (c+1)=e^{i\theta}\Psi_{c}(c).
\ee
This equation reflects the periodicity condition should be imposed onto the wave function $\Psi_{c} $ of the physical Minkowskian (YM-Higgs) vacuum at shifts of the topological dynamical variable $c(t)$ on integers $n\in \bf Z$
$$ c(t)\to  c(t)+n.    $$
Herewith the quantum-mechanical s of Eq. (\ref{per}) is quite transparent: equal probabilities to detect different topologies in the Minkowskian (YM-Higgs) vacuum quantized by Dirac \cite{Dir}. 
Then the purely real energy-momentum spectrum (\ref {Pcr}),  (\ref {rspec}) of the mechanical rotator (\ref{rot}) can be read also from the periodicity constraint (\ref {per}).
Thus  the field theoretical analogy of the Josephson effect \cite{Pervush3} was got in \cite{David3} for the Minkowskian Gauss-shell non-Abelian Higgs theory quantized by Dirac \cite{Dir}. 

Coleman et al. \cite{Col} were the first who guessed an effect similar to (\ref{sem}) in QED$_{(1+1)}$, but from a classical point of view. 

The quantum
treatment of the Josephson  effect in (Minkowskian) QED$_{(1+1)}$ was discussed then in the papers \cite{Pervush3, Ilieva2,Gogilidze} \footnote{In particular, it was demonstrated in Ref. \cite{Ilieva2} that the Josephson effect in (Minkowskian) QED$_{(1+1)}$ comes to circular motions of topologically degenerated gauge fields $ A_1^{(n)}(x,t)$ around the circle $S^1$ of the infinite radius
Herewith such closed trajectories of  infinite radii is the result identifying \cite {Ilieva2}  points 
$$ A_1^{(n)}(x,t)= \exp (i \Lambda ^{(n)}(x)) (A_1 (x,t) + i \frac{\partial_1}{e}) \exp (-i \Lambda ^{(n)}(x)), \quad n \in {\bf Z},           $$ 
  in the QED$_{(1+1)}$ configuration space $\{A_1(x,t)\}$ at the spatial infinity.
Here $\Lambda ^{(n)}(x)$ are $U(1)$ gauge matrices possessing the spatial asymptotic \cite {Pervush3, Ilieva2}  
$$ \Lambda ^{(n)}(x)= \hbar ~2\pi n \frac {x} {\pm R}   
      $$
(with $R$ standing for the spatial infinity).
The immediate manifestation of the Josephson effect in QED$_{(1+1)}$ \cite{Ilieva2,Gogilidze}, involving identifying points, at the spatial infinity,
 in the field configuration $\{A_1(x,t)\}$, is the existence of the vacuum electric field
$$   G_{10}=\dot c(t) \frac {2\pi}{e} =e(\frac {\theta}{2\pi}+k),
             $$
that  again never vanishing until $\theta\neq 0$ (this equation was derived by. Coleman et al. \cite{Col})}, and we recommend our readers Refs. \cite{Pervush3, Ilieva2,Gogilidze} for a detailed study the topic "Minkowskian QED$_{(1+1)}$". 

\medskip
The explicit expressions for the rotary momentum $I$ and the momentum $P_c$ proper to the physical (YM-Higgs) vacuum quantized by Dirac \cite{Dir} obtained in the recent papers \cite{Pervush2,LP1, David3, LP2, David1} allowed to derive, in Refs.  \cite{LP1,LP2}, the form of the vacuum (Bose condensation) Hamiltonian $H_{\rm cond}$ written down over the YM Gauss law constraint (\ref{Gau}) surface 

\be
\label{Hamilton}
H_{\rm cond}= \frac {2\pi}{g^2\epsilon}[ P_c^2 (\frac {g^2}{8\pi^2})^2+1].
\ee  
This Hamiltonian contains the "electric" and "magnetic" contributions.

The "electric" contribution to the constraint-shell Bose condensation Hamiltonian (\ref {Hamilton}) \cite{LP1,LP2} is determined by the rotary action functional (\ref{rot}) (associated with vacuum "electric" monopoles (\ref{se})- (\ref{el.m})), while  the "magnetic" contribution in this Hamiltonian is 
\cite{David3}
\be 
\label{magn.e1}
\frac{1}{2}\int \limits_{\epsilon}^{\infty } d^3x [B_i ^a(\Phi_k)]^2 \equiv \frac{1}{2}V <B^2> =\frac 1{2\alpha_s}\int\limits_{\epsilon}^{\infty}\frac {dr}{r^2}\sim \frac 1 2  \frac 1
{\alpha_s\epsilon}= 2\pi \frac{gm}{g^2\sqrt{\lambda}}=\frac {2\pi} {g^2\epsilon}. 
\ee 
This "magnetic" contribution is associated with the Bogomol'nyi equation (\ref {Bog}).

The remarkable feature  of the constraint-shell Bose condensation Hamiltonian (\ref {Hamilton}) is its manifest Poincar$\grave{}e$ (in particular, CP) invariance stipulated by the topologically momentum squared, $ P_c^2$, entering this Hamiltonian. 

This result for the Bose condensation Hamiltonian (\ref {Hamilton}) in the Minkowskian non-Abelian Higgs model quantized by Dirac \cite{Dir} is an alternative to the so-called $\theta$-term \cite{Cheng} arising in the effective Lagrangian of the Euclidian instanton non-Abelian theory \cite{Bel},
\be \label{eL}
{\cal L}_{\rm eff} = {\cal L}+ \frac{g^2\theta}{16\pi^2} ~{\rm tr}~ ( F_{\mu \nu}^a \tilde F^{\mu \nu}).
\ee
This effective Lagrangian of the Euclidian instanton non-Abelian theory \cite{Bel} is directly proportional to the pseudomentum $\theta$, and this determines the  Poincar$\grave{}e$ (CP) covariance of the Euclidian instanton non-Abelian theory, worsening its renormalization properties.

 This Poincar$\grave{}e$ (CP) covariance of the Euclidian instanton effective Lagrangian (\ref{eL}) \cite{Cheng} is the essence of the instanton CP-problem and may be avoided in the Minkowskian non-Abelian Higgs model quantized by Dirac (as we see this with the example of the Poincar$\grave{}e$ invariant Bose condensation Hamiltonian (\ref {Hamilton}) \cite{LP1,LP2} of that model).

\medskip
Generally speaking, the manifest  Poincar$\grave{}e$ invariance of the constraint-shell vacuum Hamiltonian (\ref {Hamilton}) in the Minkowskian non-Abelian Higgs model quantized by Dirac \cite{Dir} is somewhat a paradoxical thing in the light of the S (relativistic) covariance \cite{Pervush2} (\ref {ltf}), (\ref {ltf1}) of topological Dirac variables (\ref {per.Diraca}).

Indeed, the manifest Poincar$\grave{}e$ invariance of the constraint-shell vacuum Hamiltonian (\ref {Hamilton}) is due absorbing \cite{Cheng}  Gribov topological multipliers $v^{T(n)}({\bf x})$ in the G-invariant YM tension tensor squared $(F_{\mu \nu}^a)^2$.

\bigskip
C. {\it Rising "golden section" potential of the instantaneous interaction}. 

As  demonstrated in Refs. \cite{Pervush2,LP1, David3, LP2, David1}, in the YM BPS monopole background (turning into the Wu-Yang monopole background \cite{Wu} at the spatial infinity), the Green function $G_{ab}({\bf x}, {\bf y})$ entering the "potential" item (\ref {cint}) \cite{Werner,David1}
in the constraint-shell Hamiltonian of the Minkowskian non-Abelian Higgs model (quantized by Dirac \cite{Dir}) may be decomposed into the complete set of  orthogonal
vectors in the colours space
\be
\label{complete set}
G^{ab}({\bf x},{\bf y})= [n^a(x) n^b(y)V_0(z)+ \sum \sb {\alpha=1,2} 
e^a_ \alpha (x)e^{b\alpha}(y)V_1(z)];\quad (z=\vert {\bf x}-{\bf y }\vert).
\ee
This  equation involves two instantaneous interaction potentials: $ V_0(z)$ and $ V_1(z)$.

The first of these potentials, $ V_0(z)$, proves to be the Coulomb  type potential
\be
\label{Cp}
V_0 (\vert {\bf x}- {\bf y} \vert) = 
-1/ 4\pi ~\vert {\bf x}- {\bf y} \vert ^{-1} + c_0,
\ee
where $c_0$ is a constant. 

The second potential, $ V_1(z)$, is the so-called "golden section" potential
\be
\label{ris1}
V_1 (\vert {\bf x}-{\bf y} \vert)=
-d_1\vert {\bf x}-{\bf y} \vert ^{-1.618}+c_1\vert {\bf x}-{\bf y} \vert^{0.618},
\ee
involving constants $d_1$ and $c_1$ \footnote{Specifying constants $d_1$, $c_0$ and $c_1$, entering the  potentials $V_1$ and $V_0$, respectively, is, indeed, a very important thing.
These constants can depend, for instance, on a flavours mass scale $m_f$ or the temperature $T$ of surroundings about the system of quantum fields that the investigated Minkowskian non-Abelian Higgs model includes.
The author is grateful personally to Prof. D. Ebert who has drawn his attention to the necessity to select correctly constants entering expressions for instantaneous interaction potentials (this was during L. L. visit of Alexander von Humboldt  University Berlin in August 2002).}. \par 

The "golden section" potential (\ref{ris1})
 (unlike  the Coulomb-type one, (\ref{Cp})) implies the rearrangement of the naive perturbation series and the spontaneous breakdown of the
chiral symmetry. 
In turn, this involves the  constituent gluonic mass  in the Feynman diagrams: this mass changes
the asymptotic freedom  formula  \cite{Gr} in the region 
of  low transferred  momenta. Thus the
coupling constant $\alpha _{QCD}(q^2\sim 0)$ becomes finite.  
The  "golden section" potential (\ref{ris1}) can be also considered
 as an origin of  "hadronization" of  quarks and
gluons in QCD \cite {Pervush2, Werner, Bogolubskaja,Yura2}.\par

\bigskip
D. {\it Solving the $U(1)$-problem}.

The Dirac fundamental quantization \cite{Dir} of the Minkowskian non-Abelian Higgs model may be adapted to solving the $U(1)$-problem, i.e. finding the $\eta'$-meson mass near to modern experimental data \footnote{Modern experimental data for the $\eta'$-meson mass result $m_{\eta'}\sim 957,57$ MeV (see, e.g. the reference book \cite{Kuzm}).}.

As  demonstrated in the recent papers  \cite{Pervush2,LP1, David3, LP2, David1}, the way to solve the $U(1)$-problem in the Minkowskian non-Abelian Higgs model quantized by Dirac is associated with the manifest rotary properties of the appropriate physical vacuum involving YM and Higgs BPS monopole solutions.
The principal result obtained in the works \cite{Pervush2,LP1, David3, LP2, David1} regarding the solving of the $U(1)$-problem in the Minkowskian non-Abelian Higgs model quantized by Dirac is the following.

The $\eta'$-meson mass $m_{\eta'}$ proves to be inversely proportional to  $\sqrt I$, where the rotary momentum $I$ of the physical Minkowskian (YM-Higgs) vacuum is given by Eq. (\ref{I1}) \cite{David3}:
$$ m_{\eta'}\sim 1/ \sqrt I.     $$
More precisely,
\be
\label{mQCD}
m_{\eta'}^2 \sim \frac{C_{\eta}^2}{I V}= \frac{N_f^2}{F_{\pi}^2}\frac{\alpha_s^2<B^2>}{2\pi^3},
\ee
involving a constant $ C_{\eta}= (N_f / F_{\pi}) \sqrt {2/\pi}$, with $ F_{\pi}$ being the pionic decay constant  and $N_f$ being the number of flavours in the considered Minkowskian non-Abelian Higgs model.

The explicit value (\ref{I1}) of the rotary momentum $I$ of the physical Minkowskian (YM-Higgs) vacuum was substituted in  this equation for the $\eta'$-meson mass $m_{\eta'}$.
The result (\ref{mQCD}) for the $\eta'$-meson mass $m_{\eta'}$ is given in Refs. \cite{Pervush2,LP1, David3, LP2, David1} for the Minkowskian non-Abelian Higgs model quantized by Dirac \cite{Dir} and implemented vacuum BPS monopole solutions allows to estimate the vacuum expectation value of
  the appropriate "magnetic" field $\bf B$ (specified in that case via the Bogomol'nyi equation (\ref{Bog})) 
\be
\label{Bav} 
 <B^2>=\frac{2\pi^3F_{\pi}^2
 m^2_{\eta '}}{N_f^2\alpha_s^2}=\frac{0.06 GeV^4}{\alpha_s^2}
 \ee 
by
using  estimated $\alpha_{s}(q^2 \sim 0)\sim 0.24$  
\cite{David3,Bogolubskaja}. \par

One can assert, analysing these results obtained in Refs.  \cite{Pervush2,LP1, David3, LP2, David1} concerning specifying the $\eta'$-meson mass and estimating the vacuum "magnetic" field $\bf B$ with \linebreak $<B^2>\neq 0$, that alone going over to the Dirac fundamental quantization scheme \cite{Dir} from the "heuristic" one \cite{FP1} when considering the Minkowskian non-Abelian Higgs model is quite justified by these realistic results near to modern experimental data (for instance, \cite{Kuzm}).

In particular, the crucial role of collective solid rotations (\ref{rot}),  (\ref{I1}) \cite{David3} inside the physical Minkowskian (YM-Higgs) vacuum (they are the direct display of the Dirac fundamental quantization \cite{Dir}  of the Minkowskian non-Abelian Higgs theory) in Eq.  (\ref{mQCD}) for the $\eta'$-meson mass and Eq.  (\ref{Bav}) for  $<B^2>$ is highly transparent and impressing \footnote{ It is worth to recall here two alternative "answers"  to the question about the mass of the $\eta'$-meson that were given basing on the Euclidian non-Abelian theory \cite{Bel} involving instantons. \par 
It is, firstly, the "massless variant"  given in the paper \cite{Suskind}. 
This variant was associated with  maintaining the $\theta$-angle dependence in the effective Lagrangian ${\cal L}_{\rm eff}$ \cite{Cheng}, (\ref {eL}), in the Euclidian non-Abelian instanton QCD. \par
In this case the $\theta$-angle is covariant under chiral rotations \cite{Cheng}
$$ \theta \to  \theta '= e^{i\alpha Q_5}\theta$$
(involving  the axial charge $ Q_5=\psi\gamma_5 \bar \psi$ and an arbitrary parameter $ \alpha $), and small oscillations around the given  $\theta$-angle corresponds to a massless and unphysical fermion implying the \it Kogut-Suskind pole \rm \cite{Suskind} in the appropriate propagator. 
The diametrically opposite answer, in comparison with \cite{Suskind}, to the question about the mass of the $\eta'$-meson was given  in the paper \cite{Witten}, resting on the analysis of  \it planar diagrams \rm for the
strong interaction, in turn worked out in the paper \cite{Hooft2}, and the ABJ-anomalies  
theory \cite{Cheng,ABJ}. \par
The principal idea of the work \cite{Witten} was deleting the $\theta$-angle dependence from the effective QCD Lagrangian in the Euclidian non-Abelian instanton theory \cite{Bel}. \par
As a result, the nonzero mass of the $\eta'$-meson was obtained in the work \cite{Witten}. 
This was 
one of early approaches to  solving  the $U(1)$-problem at which 
arguments in favour of the  mesonic mass  were given. \par
Unfortunately, general shortcomings  of the Euclidian non-Abelian instanton theory \cite{Bel} (for instance, the complex energy-momentum spectrum $P_{\cal N}$ \cite{Arsen, Pervush1, Galperin,rem3}, (\ref{tm}), at the zero eigenvalue $\epsilon=0$ of the $\theta$-vacuum energy) turn the   Euclidian methods \cite{Suskind, Witten} to specify the $\eta'$-meson mass into  little effective ones.  This forced to search after another ways to construct mesonic bound states than ones \cite{Suskind, Witten} proposed in the Euclidian 
non-Abelian theory \cite{Bel}. \par

In Refs.  \cite{Pervush2,LP1, David3, LP2, David1}, just such "another way" to  solve  the $U(1)$-problem, based on the  Minkowskian non-Abelian Higgs model quantized by Dirac \cite{Dir} and involving the vacuum BPS monopole solutions was proposed.}.

\bigskip 
E. {\it Fermionic rotary degrees of freedom in the Wu-Yang  monopole background}.

A good analysis of the question about the place of fermionic (quark) degrees of freedom in the  Minkowskian non-Abelian Higgs model quantized by Dirac \cite{Dir} was performed in  the recent papers \cite{David3, David1}.

For instance, as we have seen above, G-invariant fermionic currents \cite{Pervush2}
\be \label{fc} j_\mu^{Ia}= g\bar \psi^I (\lambda ^a /2) \gamma_\mu \psi^I,\ee 
belonging (as defined in Ref. \cite{David1})  to the excitation spectrum over the physical Minkowskian (YM-Higgs) vacuum involving the vacuum BPS monopole solutions, enter total G-invariant currents (\ref{tocar}) \cite{David1}, satisfying the Gauss law constraint (\ref {cur-t}) \cite{LP1,David1}. 

\medskip
New interesting properties acquire  fermionic (quark) degrees of freedom  $\psi^I$, $\bar \psi^I$ in Minkowskian constraint-shell QCD involving the spontaneous breakdown of the initial $SU(3)_{\rm col}$ gauge symmetry in the
\be \label{break} SU(3)_{\rm col} \to SU(2)_{\rm col} \to U(1) \ee
way.
Actually, such Minkowskian constraint-shell QCD is the particular case of the \linebreak Minkowskian non-Abelian Higgs models quantized by Dirac \cite{Dir}.

The only specific of  Minkowskian constraint-shell QCD (in comparison with the constraint-shell Minkowskian (YM-Higgs) theory) is the presence therein of three Gell-Mann matrices $\lambda^a$, generators of $ SU(2)_{\rm col}$ (just these matrices would enter G-invariant quark currents $ j_\mu^{Ia}$ in of  Minkowskian constraint-shell QCD). 
In the constraint-shell Minkowskian (YM-Higgs) theory, involving the initial $ SU(2)$ gauge symmetry, the Pauli matrices $\tau^a$ ($a=1,2,3$) would replace the Gell-Mann $\lambda^a$ ones.

The very interesting situation, implying lot of important consequences, takes place to be in Minkowskian constraint-shell QCD involving the spontaneous breakdown (\ref{break}) of the initial $SU(3)_{\rm col}$ gauge symmetry  when the antisymmetric 
Gell-Mann matrices 
\be \label{choice} \lambda_2,\lambda_5, \lambda _7 \ee  
are chosen to be the generators of the
$SU(2)_{\rm col}$ subgroup in (\ref{break}), as it was done in Refs. \cite{Pervush2, David3, David1}. 

As  demonstrated in Ref. \cite{David3}, the "magnetic" vacuum field $ B^{i a}(\Phi_i)$ corresponding to Wu-Yang monopoles $\Phi_i $ \cite{Wu} acquires the form
\be \label{b11}
b_i^a=\frac 1 g \epsilon_{iak}\frac{n_k(\Omega)}{ r}; \quad
n_k(\Omega)=\frac{x^l\Omega_{lk}}{r}, \quad n_k(\Omega) n^k(\Omega)=1;
\ee 
in terms of  the antisymmetric 
Gell-Mann matrices $\lambda_2, \lambda_5, \lambda_7$, 
(\ref{choice}), with $\Omega_{lk}$ being an orthogonal matrix in  the colour space.

For the "antisymmetric" choice (\ref{choice}), we have
\be \label{b12} b_i\equiv\frac{g}{2i}  b_{ia} \tau  ^a=
g\frac{b_i^1\lambda^2+b_i^2\lambda^5+b_i^3\lambda^7}{2i};~~~~~ b_i^a=\frac{\epsilon^{aik}n^k}{gr} \quad ( \tau_1 \equiv \lambda_2,\tau_2 \equiv \lambda_5,\tau_3 \equiv\lambda _7). \ee 

\medskip
On the other hand, the important task that Minkowskian constraint-shell QCD is called to solve is getting spectra of mesonic and baryonic bound states.
As we   have noted in Section 2, the presence of such hadronic  bound states in a gauge model violates the gauge equivalence theorem \cite{Taylor, Slavnov1, Arsen}.
As in the case of collective vacuum excitations, this implies the identity (\ref{spur2}), involving spurious Feynman diagrams (SD).

A detailed analysis how to apply the Dirac fundamental quantization method \cite{Dir} to constructing hadronic  bound states was performed in the papers \cite{Yura3, Pysinin}, and then such analysis was repeated in Ref. \cite{Pervush2}.
The base of the approach to constructing hadronic  bound states that was proposed in \cite{Pervush2,Yura3, Pysinin} is the so-called {\it Markov-Yukawa prescription} \cite{Markov-Yukawa}, the essence of which is \cite{Pervush2, Markov-Yukawa} in separating {\it absolute}, $X_\mu=(x+y)_\mu/2$, and {\it relative}, $z_\mu=(x-y)_\mu$, coordinates, involving treatment of (mesonic) bound states as {\it bilocal fields}
\begin{eqnarray} \label{3-1}
 {\cal M}(x,y) = e^{iMX_0} \psi(z_i) \delta(z_0).
 \end{eqnarray}
The important feature of such bilocal fields is observing two particles (say, some quark $q$ and antiquark $\bar q$) as a bound state at one and the same time. 

This principle of the simultaneity has more profound mathematical meaning \cite{Pervush2,Yura2} as the
constraint of irreducible nonlocal representations of the Poincare group for arbitrary bilocal field $ {\cal M}(x,y) \equiv {\cal M}(z \vert X)$:
\begin{eqnarray}\label{3-3}
 z_\mu \frac{\partial}{\partial X_\mu} {\cal M}(z \vert X) = 0,
 ~~~~~~~~~~~~~~{\cal M}(z|X)\equiv {\cal M}(x,y).
\end{eqnarray}
This constraint is not connected with the dynamics of interaction and realized the \linebreak Eddington
simultaneity \footnote{"A proton yesterday and electron today do not make an atom" \cite{Eddington}.}.

Thus  the constraint  (\ref{3-3})  results in the  choice of the bound state  relative coordinates  $z_\mu$  to be orthogonal to its total momentum
$  {\cal P}_\mu\equiv -i \frac \partial {\partial X_\mu}$ 
\begin{eqnarray}  \label{3-5}
(z^{\perp})_\mu = z_\mu - {\cal P}_{\mu}
 \bigl( \frac{{\cal P} \cdot z}{{\cal P}^2 } \bigr).  
 \end{eqnarray}
Moreover, at the point of the forming of the bound state with the definite total momentum ${\cal P}_\mu$, it is possible to choose the time  axis $\eta_\mu=(1,0,0,0)$ to be parallel to this total momentum: $\eta_\mu \sim {\cal P}_{\mu}$.

Therefore, \cite{Pervush2}
\begin{eqnarray} \label{3-6}
 \eta_\mu  {\cal M} (z \vert X)   \sim {\cal P}_{A\mu} {\cal M} (z
 \vert X) = \frac{1}{i} \frac{\partial}{\partial X_\mu} {\cal M} (X
 \vert z).
 \end{eqnarray}
In the rest reference frame $\eta_\mu $ chosen in the (\ref{3-6}) way, the instantaneous interaction between the particles forming the given bilocal bound state ${\cal M} (X
 \vert z)$ takes the form \cite{Pervush2}
\begin{eqnarray} 
\label{rc1} 
 { W}_{I} = \int d^4 x d^4 y \frac{1}{2} j_{\eta}^{D}(x) V_I(z^{\perp})  j_{\eta}^{D}(y) \delta(\eta \cdot z).  
\end{eqnarray} 
 This equation involves manifestly G-invariant and S-covariant fermionic currents 
$$  j_{\eta}^{D} = e \bar {\psi}^D \rlap/\eta \psi^D; \quad \rlap/\eta \equiv \eta_\mu \gamma ^\mu  $$
 (attached to the rest reference frame $\eta_\mu $ chosen in the (\ref{3-6}) way and implicating fermionic Dirac variables $\psi^D$, $\bar\psi^D$).
$ V_I(z^{\perp})  $ is the instantaneous interaction potential between the particles forming the bilocal bound state ${\cal M} (X
 \vert z)$. 
The manifest S-covariance of the constraint-shell action functional (\ref{rc1}) follows immediately from the transformation law (\ref {ltf1}) \cite{Pervush2} for fermionic Dirac variables $\psi^D$, $\bar\psi^D$.

 Incidentally, note that upon extracting G-invariant fermionic currents $j_\mu^{aI}$ (\ref{fc}) from the total ones \cite{David1} (\ref {tocar}) it is possible to write down the constraint-shell action functional of the (\ref{rc1}) type for the Minkowskian Higgs model with vacuum BPS monopole solutions describing the instantaneous interaction between these fermionic currents, attached to the rest reference frame $\eta_\mu $ (\ref{3-6}) and  involving herewith the Green function $G_{ab}({\bf x},{\bf y})$ of the Gauss law constraint (\ref {cur-t}) \cite{LP1,LP2,David1}.
  It may, in turn, be decomposed in the (\ref {complete set}) way, implicating the Coulomb  type potential $V_0(z)$, (\ref {Cp}), and the "golden section" one, $V_1(z)$, (\ref {ris1}).

\medskip
In the papers \cite{Pervush2,Yura3, Pysinin} the algorithm is given for the derivation of    mesonic  bound states spectra utilizing the Markov-Yukawa prescription \cite{Markov-Yukawa}, us outlined above.
Omitting details of this algorithm and referring our readers to Ref. \cite{Pervush2,Yura3, Pysinin} (with the literature cited therein) for the detailed acquaintance with the question, now note that the important step of this algorithm is solving of the Dirac  equation for a fermion (quark) in the BPS (Wu-Yang) monopole background.

For the spontaneous breakdown of the initial $SU(3)_{\rm col}$ gauge symmetry in the (\ref {break}) way, involving herewith antisymmetric Gell-Mann matrices $\lambda_2$, $\lambda_5$, $\lambda_7$ as generators of the "intermediate" $SU(2)_{\rm col}$ gauge symmetry, this BPS (Wu-Yang) monopole background takes the look (\ref {b11}) \cite{David3}.
To  write down the Dirac equation for a quark in the BPS (Wu-Yang) monopole background, note that 
each fermionic (quark) field may be decomposed by the complete set of the generators 
of the  Lee group $SU(2)_{col}$
(i.e. $\lambda_2, \lambda_5, \lambda_7$ in the considered case) completed by
the unit matrix $\bf 1$ \cite{Cheng}.
This involves the following decomposition \cite{David3} of a quark field by the antisymmetric Gell-Mann matrices $\lambda_2$, $\lambda_5$, $\lambda_7$
\be \label{decompoz}
\psi_{\pm}^{\alpha,\beta} =s_{\pm} \delta^{\alpha,\beta} + v_{\pm}^j\tau_j^{\alpha,\beta},
\ee
involving some $SU(2)_{\rm col}$ isoscalar,  $s_{\pm}$, and isovector, $v_{\pm}$,
amplitudes.
 $+,-$ are spinor indices,  $\alpha,\beta$ are 
 $SU(2)_{\rm col}$ group space indices and 
$$(\lambda_2, \lambda_5, \lambda_7)\equiv (\tau_1,\tau_2,\tau_3).$$
The mix of group and spinor indices generated by Eqs. (\ref{b11}), (\ref{b12}) for the BPS (Wu-Yang) monopole background allows  then to derive, utilising the decomposition (\ref{decompoz}),
the system of differential equations in partial derivatives \cite{David3}
\be \label{Desyst} (\mp q_o+m) s_{\mp} {\mp}i(\partial_a + \frac{n_a}{r}) v^a_{\pm}=0;
\ee \be \label{Desyst1} (\mp q_o+m) v^a_{\mp} {\mp}i(\partial^a - \frac{n^a}{r}) s_{\pm}
-i\epsilon^{jab}\partial_jv_{\pm}^b =0
\ee 
(implicating the mass $m$ of a quark and its complete energy $q_0$), mathematically equivalent to the Dirac equation
\be
\label{De}
 i\gamma_0 \partial_0 \psi + \gamma_j [i\partial_j \psi+ \frac{1}{2r} \tau_a \epsilon^{ajl}n_l \psi] -m \psi= 0 \ee 
for a quark in the BPS (Wu-Yang) monopole background. 

The decomposition (\ref{decompoz}) \cite{David3} of a quark field implies \cite {Gelfand} that $v_{\pm}^j\tau_j^{\alpha,\beta}$ is a  three-dimensional axial vector in the colour space. 
Thus the spinor (quark) field $\psi_{\pm}^{\alpha,\beta}$ is transformed, with the "antisymmetric"
choice
$\lambda_2, \lambda_5, \lambda_7$,
by the \it reducible \rm
representation of the $SU(2)_{col}$ group that is the direct sum of the
identical representation $\bf 1$ and  three-dimensional  axial vector representation,  we denote as ${\bf 3}_{ax}$.

A  new situation, in comparison  with the
usual $SU(3)_{\rm col}$ theory in the Euclidian space $E_4$ \cite{Cheng}, appears in 
this case. 
That theory was worked out by Greenberg 
\cite{Greenberg}  Han and Nambu  \cite{HN,Nambu}; its goal was getting  hadronic wave
functions  (describing bound quark states) with the correct spin-statistic connection. 
To achieve this, the \it irreducible \rm colour triplet (i.e.  three additional 
degrees of freedom of quark  colours\rm, forming the {\it polar} vector in  the $SU(3)_{\rm col}$ group space), was introduced. There was postulated that  only  colour singlets are physical
observable states. So the task of the colours confinement was outlined.\par 
The transition  to the Minkowski space in Minkowskian constraint-shell QCD quantized by Dirac \cite{Dir} and involving the (\ref {break}) breakdown of the $SU(3)_{col}$ gauge symmetry, the antisymmetric   Gell-Mann matrices $\lambda_2$, $\lambda_5$, $\lambda_7$ and BPS (Wu-Yang) physical background, allows to introduce the new, reducible,  representation of the $SU(2)_{\rm col}$ group with axial colour vector and colour scalar. \par 
In this situation the question about the physical sense of the axial colour vector $v_{\pm}^j\tau_j^{\alpha,\beta}$ is posed. \par 
For instance, it may be assumed that the axial colour vector $v_{\pm}^j\tau_j^{\alpha,\beta}$ has the form ${\bf v}_1= {\bf r} \times\bf K$, with $\bf K$ being the polar colour vector 
($SU(2)_{\rm col}$ triplet).
These quark rotary degrees of freedom corresponds to rotations of fermions together with the gluonic BPS monopole vacuum describing by  the free rotator action (\ref{rot}) \cite{David3}. The latter one is induced by  vacuum "electric" monopoles
(\ref{el.m}). 
These vacuum "electric" fields are, apparently, the cause of above  fermionic rotary degrees of freedom \rm (similar to rotary singlet terms in two-atomic molecules;  see e.g. \S 82 in \cite{Landau3})  \footnote{ A good analysis of the Dirac system (\ref{Desyst}), (\ref{Desyst1}) for 
isospinor fermionic fields (in the YM theory)  in the background field of a
(BPS, Wu-Yang) monopole was carried out in  the work  \cite{Rebi}.  
In  that work   was obtained the Dirac system alike (\ref{Desyst}), (\ref{Desyst1}) \cite{David3}, and also by means of the decomposition of a fermionic field by the $SU(2)$ generators, the Pauli matrices.}.

More exactly, repeating the arguments of Ref. \cite{Pervush3}, one can "nominate" the candidature of the "interference item"
\be \label{ii}
\sim Z^a j_{Ia0}
\ee
in the constraint-shell Lagrangian density of Minkowskian QCD quantized by Dirac \cite{Dir} between the zero mode solution $ Z^a $ to the Gauss law constraint (\ref {Gau}) (involving vacuum "electric" monopoles (\ref {el.m}), generating the rotary action functional $W_{\rm coop}$, (\ref {rot}), for the physical Minkowskian non-Abelian BPS monopole vacuum) and the G-invariant quark current $ j_{Ia0}$ \cite{Pervush2} (\ref {fc})  belonging to the excitation spectrum over this physical vacuum, as the source of fermionic rotary degrees of freedom ${\bf v}_1$.

\medskip
 The appearance of fermionic rotary degrees of freedom ${\bf v}_1$ in Minkowskian constraint-shell QCD quantized by Dirac \cite{Dir} confirms indirectly the existence of the BPS monopole background in that model (coming to the Wu-Yang one \cite{Wu} at the spatial infinity).
These fermionic rotary degrees of freedom testify in favour of nontrivial topological collective vacuum dynamics proper  to the Dirac fundamental quantization \cite{Dir} of Minkowskian constraint-shell QCD (this vacuum dynamics was us described above).
\section{Discussion}
  First of all note that the experimental detection of fermionic rotary degrees of freedom ${\bf v}_1$ as well as the "golden section" instantaneous interaction potential $V_1(z)$, (\ref {ris1}), between quarks, can be a good confirmation of the Dirac fundamental quantization \cite{Dir} of Minkowskian constraint-shell QCD involving physical BPS monopole vacuum possessing manifest superfluidity and various rotary effects.
This should be  equally valid as the results  obtained in Refs. \cite{Pervush2,LP1, David3, LP2, David1} concerning obtaining the $\eta'$-meson mass $m_{\eta'}$ (\ref {mQCD}).

\medskip
The ``theoretical plan'' for further  development of the Minkowskian Higgs model quantized by Dirac \cite{Dir} may be associated, in the first place, with the following assumption called to explain the nontrivial topological collective vacuum dynamics inherent in that model.
This is the assumption \cite{disc}  about the "discrete group geometry" for the initial (say, $SU(2)$) and residual (say, $U(1)$) gauge groups in the Minkowskian Higgs model.
This assumption was made already in the work \cite{Pervush1}.

Ibid  there was demonstrated that a gauge group $G$, involving (smooth) stationary transformations, say
\be \label{gaug}
A '_\mu ({\bf x}, t) = v^{-1} ({\bf x})
A _\mu ({\bf x}, t) v ({\bf x})+ v({\bf x}) \partial _\mu v^{-1}({\bf x}),
\ee
may  always be factorised in the "discrete" way as
\be
\label{fact} 
G\simeq G_0\otimes {\bf Z} \equiv {\widetilde G}; \quad {\bf Z}=G/G_0.    
\ee 
 Note that, in difinition, $\pi_1 (G_0)=0$, i.e. $G_0$ is \cite{Al.S.} the  {\it one-connected and topologically trivial} component  in the generic $\widetilde G$ gauge group factorised  in the (\ref{fact}) way. \par
Moreover, $G_0$ is the \it maximal connected component \rm in $G$ (in the terminology  \S \S ~T17, T20 in \cite{Al.S.}): $\pi_0(G_0)=0$. \par 

That  implies \cite{Al.S.} 
$$ \pi_0 [G_0 \otimes {\bf Z}]= \pi_0 ({\bf Z})= {\bf Z}.$$
The latter relation indicates transparently the discrete nature of the ${\widetilde G}$ group space. \par
More exactly, the  ${\widetilde G}$ group space consists of different topological sectors (each with its proper topological number $n\in {\bf Z}$), separated by {\it domain walls}.
Additionally, the factorisation (\ref{fact}) reflects the essence of Gribov topological \rm "copying" \cite{Gribov} of "small" gauge transformations.  
\par
\medskip
On the other hand, since (\ref {fact}) is   only an isomorphism, there is a definite freedom in assuming that the gauge group $G$ possesses a "continuous" or "discrete" geometry. 
In particular, in the Euclidian non-Abelian model \cite{Bel} involving instantons,  the "continuous" geometry should be assumed for the $SU(2)$ group space. 
It is associated with the absence of any nonzero mass scale in this model.
The thing is that domain walls between different topological sectors would become infinitely wide in this case.
\par
Generally speaking, the width of a  domain wall is roughly proportional to the inverse of the lowest mass of all  physical particles being present in the (gauge) model considered \cite{Ph.tr}. 
Thus domain walls are really infinite in the Euclidian instanton model \cite{Bel}.
In  this case any transitions \cite{Cheng} are impossible between vacua (say, $\vert n>$ and \linebreak $\vert n+1>$) with different topological numbers since latter ones belong to topological domains separated by infinitely wide walls. 

\medskip
In principle different situation is in the Minkowskian Higgs model quantized by Dirac \cite{Dir} and involving vacuum gauge and Higgs BPS monopole solutions.
In this model a natural  mass scale may be introduced.
For instance, the effective Higgs mass $m/\sqrt\lambda$ may be treated as  a mass scale, depending indeed on the distance $r$ via Eq. (\ref{masa}) (because $V\sim r^3$). 
This  creates objective prerequisites for utilizing the "discrete" representations of the ${\widetilde G}$ type \cite{Pervush1}, (\ref{fact}), for the initial, $SU(2)$, and residual, $U(1)$, gauge symmetries groups in the Minkowskian Higgs model quantized by Dirac \cite{disc}
\be
 \label{fact2}
{ \widetilde SU}(2)\simeq G_0\otimes {\bf Z}; \quad    { \widetilde U}(1) \simeq U_0 \otimes {\bf Z},
 \ee
respectively.

As a result, the {\it degeneration space} ({\it  vacuum manifold}) $$R_{YM}\equiv SU(2)/U(1)$$
 in this Minkowskian Higgs model acquires the "discrete" form 
\be
 \label{RYM}
 R_{YM}= {\bf Z}\otimes G_0/U_0. \ee
Obviously, $ R_{YM}$ is the discrete space consisting of topological domains separated by domain walls.

If the Minkowskian Higgs model quantized by Dirac \cite{Dir} involves vacuum gauge and Higgs BPS monopole solutions, in this case it is quite naturally to suppose that the typical wide of such domain walls is $\epsilon(r)$, with $\epsilon(r)\sim (m/\sqrt\lambda)^{-1}(r) $ given by Eq. (\ref{masa}).

From Eq. (\ref{masa}) it can be concluded \cite{disc} that  $\epsilon$ disappears in the infinite  spatial volume limit $V\to\infty$, while it is maximal at the origin of coordinates (herewith it can be set $ \epsilon (0)\to\infty$). This means, due to the  reasoning \cite{Ph.tr}, that walls between topological domains inside $R_{\rm YM}$ become infinitely wide, $O(\epsilon (0))\to\infty$, at the origin of coordinates. The fact $\epsilon(\infty)\to 0$ is  also meaningful.  This implies  actual merging of topological domains inside the vacuum manifold $R_{\rm YM}$, (\ref{RYM}), at the spatial infinity. This merging of topological domains promotes the infrared topological confinement (destructive interference) of Gribov "large" multipliers $ v^{(n)}({\bf x})$ in gluonic and quark Green functions in all the orders of the perturbation theory (in the spirit of Ref. \cite{Azimov}).
 On the other hand, It becomes obvious that the effective Higgs mass $m/\sqrt\lambda$ (as the value inversely proportional to $\epsilon$) is really can be treated as a Wegner variable, disappearing in the ultraviolet fixed point (i.e. at the origin of coordinates) \cite{Kadanoff}. 

\medskip
It may be demonstrated that the vacuum manifold $ R_{YM}$ in the Minkowskian Higgs model quantized by Dirac \cite{Dir} in its "discrete" representation (\ref {RYM}) possesses three kinds of topological defects.
The first kind of topological defects are domain walls between different topological sectors of that Minkowskian Higgs model, us discussed above.
The criterion of domain walls existing in a (gauge) model is a nonzero (for example, infinite) number $\pi_0$ of connection components in the appropriate degeneration space.
In particular,
$$  \pi_0 (R_{YM})={\bf Z}  $$
because of (\ref {fact2}).

\medskip
The next kind of topological defects inside the discrete YM vacuum manifold $ R_{YM}$ are {\it point hedgehog topological defects}.
This type  of topological defects comes to the vacuum "magnetic" field $\bf B$, generated by the Bogomol'myi equation (\ref{Bog}), singular at the origin of coordinates, as it was shown in Ref. \cite{BPS}.
Actually, $\vert{\bf B\vert }\sim O(r^{-2})$.
From the topological viewpoint, the criterion for point (hedgehog) topological defects to exist in a (gauge) theory is  the nontrivial group $\pi_2$ of two-dimensional ways for the appropriate degeneration space (vacuum manifold).

Moreover, denoting as $G$ the initial gauge symmetry group in the considered model and as $H$ the residual one (then $R=G/H$  will be the vacuum manifold in that model), it may be proved \cite{Al.S.} that always
$$  \pi_2 R=\pi_1 H$$
(with $\pi_1 H $ being the fundamental group of one-dimensional ways in $H$), and herewith if $\pi_1 H \neq 0$,  point (hedgehog) topological defects exist in a (gauge) theory \cite{Al.S.}. 

In particular, the topological relation 
\be \label {point1}
\pi_2(R_{YM})=\pi_1 \widetilde U(1)={\bf Z} 
\ee
is the criterion of point (hedgehog) topological defects in the Minkowskian Higgs model quantized by Dirac \cite{Dir}. 

Geometrically, point topological defects are concentrated in a coordinate region topologically equivalent to a two-sphere $S^2$ (in particular, point hedgehog topological defects are always
concentrated in  a two-sphere with its centre lying in the origin of coordinates).
Just in such coordinate regions the thermodynamic equilibrium (at   a Curie point $T_c$ in which the appropriate second-order phase transition occurs) corresponding to the minimum of the action functional set over the vacuum manifold $R$ in a (gauge) model is violated \cite{Al.S.}.
This violating involves singularities in order parameters.
An example of such singularities order parameters found in  (gauge) models 
with point topological defects is the $O(r^{-2})$ behaviour \cite{BPS} of the vacuum "magnetic" field $\bf B$ in the Minkowskian Higgs model involving vacuum BPS monopole solutions.

\medskip
In conclusion, the vacuum manifold $ R_{YM}$ contains the third kind of topological defects, {\it thread topological defects}.
The criterion for thread topological defects to exist in a (gauge) theory is  the topological relation \cite{Al.S.}
\be \label{tr.def.}
\pi_1 R= \pi_0 H\neq 0.
\ee
In particular, 
\be
\label{nit'}
\pi_1 (R_{YM})= \pi_0 ~\widetilde U(1)={\bf Z}.
\ee
Thus thread topological defects exist in the Minkowskian Higgs model quantized by Dirac \cite{Dir} (implicating vacuum BPS monopole solutions) in which the "discrete" vacuum geometry (\ref{RYM}) is assumed.  

Geometrically, thread topological defects cause violation of the thermodynamic equilibrium along definite lines (for instance, rectilinear ones) in the given vacuum manifold.
It may be demonstrated, repeating thc arguments of Ref. \cite{Al.S.}, that thread topological defects possess the manifest $S^1$ topology (for instance,  for "rectilinear" thread topological defects it is highly transparent).

\medskip
Point (hedgehog) topological defects always present  in the Minkowskian Higgs model involving vacuum monopole solutions, irrelevantly to the way in which this model is quantized: either  this is the FP "heuristic" quantization scheme \cite{FP1} or the Dirac fundamental one \cite{Dir}.

Besides the BPS monopoles \cite{Al.S.,BPS,Gold} and Wu-Yang ones \cite{Wu}, granted a great attention in the present study, 't Hooft-Polyakov monopoles \cite{H-mon, Polyakov} also the very important kind of monopole solutions with which modern theoretical physics deals. 
Indeed, the analysed Minkowskian Higgs models involving vacuum monopole solutions and point (hedgehog) topological defects associated with these vacuum monopole solutions confine themselves within the FP "heuristic" quantization scheme \cite{FP1}.

On the other hand, it is sufficient to assume the "continuous", $\sim S^2$, vacuum geometry in the Minkowskian Higgs models \cite{Al.S.,BPS,Gold,Wu, H-mon, Polyakov} with monopoles in order to quantize them in the "heuristic" \cite{FP1} wise.
We have already discussed this with the example of vacuum BPS monopole solutions \cite{Al.S.,BPS,Gold} in which the Bogomol'nyi equation (\ref{Bog}) and the Bogomol'nyi bound $E_{\rm min}$, (\ref{Emin}) were derived \cite{Al.S.,rem1} just  assuming the continuous 
$$ SU(2)/U(1)\sim S^2$$
vacuum geometry and herewith without solving the YM Gauss law constraint (\ref {Gau}) \footnote{Indeed,  as it was explained in Refs. \cite{LP1,LP2,rem2}, the Bogomol'nyi equation (\ref{Bog}) is compatible with the Dirac fundamental quantization \cite{Dir} of the Minkowskian Higgs model  with BPS monopoles. 
As  discussed above, this connection between the Bogomol'nyi equation (\ref{Bog}) and the Dirac fundamental quantization of the Minkowskian Higgs model is given via the Gribov ambiguity equation (having the form (\ref {koop}), to which the Bogomol'nyi equation (\ref{Bog}) comes mathematically because of the Bianchi identity $ D~ B=0$.}.

In the analysed Minkowskian Higgs models \cite{Al.S.,BPS,Gold,Wu, H-mon, Polyakov} with monopoles there are no nontrivial (topological) dynamics, since the physical content of that models is determined by {\it stationary} vacuum monopole solutions.
Additionally,  all the "electric" tensions in the enumerated Minkowskian Higgs models are set identically in zero: $F_{0i}^a\equiv 0$.
Thus assuming the "continuous", $\sim S^2$, vacuum geometry in the Minkowskian (Higgs) models \cite{Al.S.,BPS,Gold,Wu, H-mon, Polyakov} with stationary vacuum (Higgs and YM) monopole solutions (setting additionally to zero of all  "electric" tensions) ensures quite the lawfulness
 of the "heuristic" \cite{FP1} quantization of that models \footnote{For instance \cite{rem1}, one can fix the {\it Weyl} gauge $A_0=0$ for temporal YM components in appropriate FP path integrals. This just results $F_{0i}^a\equiv 0$ if one deals with stationary monopole solutions in the analyzed Minkowskian (Higgs) models. }.

\medskip
Unlike the above discussed case, to justify the Dirac fundamental quantization \cite{Dir} of the Minkowskian Higgs model, involving the collective vacuum rotations (\ref {rot}) \cite{David1}, the "discrete" vacuum geometry of the (\ref{RYM}) type should be supposed.
More precisely, if thread "rectilinear" topological defects are contained inside the vacuum manifold $R_{YM}$, (\ref{RYM}), this can explain the discrete vacuum rotary effect (\ref {rot}) occurring in the Minkowskian Higgs model quantized by  Dirac.
Just  such rectilinear lines inside the vacuum manifold $R_{YM}$ (that are, geometrically, cylinders of effective diameters $\sim \epsilon$), localized around the axis $z$ of the chosen (rest) reference frame \cite{disc}, are associated with the Josephson effect \cite{Pervush3} in that Minkowskian Higgs model.

As we  have ascertained above, this Josephson effect comes therein \cite{David3} to the "persistent field motion" around above described rectilinear lines: with all ensuing physical consequences, including the real spectrum (\ref {Pcr}) of the appropriate topological momentum $P_c$, the never vanishing (until $\theta \neq 0$) vacuum "electric" field $(E_i^a)_{\rm min}$ \cite{David3}, 
(\ref{sem}), and the manifestly  Poincar$\grave{}e$ invariant constraint-shell Hamiltonian $H_{\rm cond}$ \cite{LP1,LP2}  (\ref {Hamilton}) of the Bose condensation.

\medskip
Investigating about the Dirac fundamental quantization \cite{Dir} of the Minkowskian Higgs model is not finished at present.
So recently, in Ref. \cite{disc}, it was  demonstrated that the first-order phase transition occurs in the constraint-shell Minkowskian Higgs model quantized by Dirac and involving vacuum BPS monopole solutions.

This first-order phase transition supplements the  second-order one always taking place in the Minkowskian Higgs model and associated with the spontaneous breakdown of the initial gauge symmetry.
The essence of the first-order phase transition occurring in the Minkowskian Higgs model quantized by Dirac and involving vacuum BPS monopole solutions is in coexisting collective vacuum rotations (described by the action functional  (\ref {rot}) \cite{David3}) and superfluid potential motions (set in the Dirac fundamental scheme \cite{Dir} by the Gribov ambiguity  equation, coming to the Bogomol'nyi  one (\ref {Bog})).

As it was demonstrated  in Ref.  \cite{Pervush1}, this first-order phase transition in the Minkowskian Higgs model quantized by Dirac comes to the claim that vacuum "magnetic" and "electric" fields: respectively, $\bf B$ and $\bf E$, are transverse
$$ D~B= D~E=0.$$
This condition, in turn, is mathematically equivalent to the system \cite{Pervush1} 
\be
\label{sistema}
E\sim D \Phi; \quad B \sim D \Phi 
\ee
of the first-order differential equations, involving  Higgs vacuum BPS monopole modes $\Phi$. 

More exactly, acting by the (covariant) derivative $D$ on the system  (\ref {sistema}), one turns the Bogomolnyi equation $ B \sim D \Phi $ (second equation in this system) into the Gribov ambiguity equation, while the first equation in (\ref {sistema}) comes then to the YM Gauss law constraint (\ref{koop}) at the constraint-shell reduction of the Minkowskian Higgs model in terms of the gauge invariant and transverse topological Dirac variables (\ref {per.Diraca}). 

Thus assuming about the "discrete" vacuum geometry of the (\ref {RYM}) type appears playing the crucial role at the above assertion that first-order phase transition occurs in the Minkowskian Higgs model quantized by Dirac \cite{Dir}, as well as at explaining other phenomena taking place in that model.

\bigskip
 The author would also like express his opinion about the further fate of gauge physics.
In author's opinion, this seems to be connected with three things.

The first one is going over to the Minkowski space (from the Euclidian $E_4$ one). This allows to avoid typical shortcomings inherent in Euclidian gauge theories (including the complex values (\ref {tm}) \cite{Arsen, Pervush1,Galperin} for the topological momentum $P_{\cal N}$ of the $\theta$-vacuum in  the Euclidian instanton model \cite{Bel}). 

The second thing is utilizing  vacuum BPS monopole solutions \cite{Al.S.,BPS,Gold} at development the Minkowskian Higgs model. 
As we have seen in the course our present discussion, this set manifest superfluid properties in that model, absent in other Minkowskian (Higgs) models with monopoles: for instance, in the Wu-Yang model \cite{Wu} or in the 't 
Hooft-Polyakov model \cite{H-mon, Polyakov}.

The third thing is the Dirac fundamental quantization \cite{Dir} of the Minkowskian Higgs model involving vacuum BPS monopole solutions, which gave perceptible results (for example, the $\eta'$-meson mass (\ref {mQCD}), near to modern experimental data, or the rising "golden section" potential (\ref {ris1}) of Refs. \cite{Pervush2,LP1, David3, LP2, David1}). 

\medskip Apart from the said, the discussed Minkowskian Higgs model quantized by Dirac (involving  vacuum BPS monopole solution BPS monopole solutions and ``discrete vacuum geometry'' (\ref {RYM}) \cite{disc}) gives the specific approach to the so-called {\it mass gap problem}. That problem was formulated as following \cite{WJ}. Experiment and computer simulations about the ``pure''YM theory without other (quantum) fields suggest the existence of a "mass gap" in the solution to the quantum versions of the YM equations. But no proof of this property is known.

In the strict mathematical language,  the   mass gap problem can be expressed in the following way.
Since the Hamiltonian $H$ of a  QFT  is the element of the Lie algebra of the Poincare 
group and the appropriate vacuum vector $\Omega$ is Poincare invariant, it is
an eigenstate with zero energy, $H\Omega=0$. The positive energy axiom (in absence of external negative potentials) asserts that in any
QFT, the spectrum of $H$ is supported in the region $[0,\infty)$.
In  this terminology,  a QFT has a {\it mass gap} if $H$ has no spectrum in the interval $[0,\Delta)$ for a $\Delta>0$ The supremum of such $\Delta$ is called the mass $m$. 
Then the YM mass gap problem can be formulated mathematically \cite{WJ} as {\it proving that for any compact simple gauge group $G$, the quantum YM theory on ${\bf R}^4$ exists and has a mass gap $\Delta>0$}. 
An important consequence of the existence of a mass gap is that for any positive constant $C<\Delta$   and for any local quantum field operator ${\cal O}(x)$  such that $\left\langle \Omega, {\cal O}\Omega\right\rangle =0$, one has
$$ \vert \left\langle\Omega, {\cal O}(x) {\cal O}(y)\Omega\right\rangle \vert \leq \exp (-C\vert x-y\vert) $$
if $\vert x-y\vert$ is sufficiently large (depending on $C$ and ${\cal O}$).

   The  Minkowskian Higgs model quantized by Dirac, here presented, contains the Higgs (and  fermionic) field modes. Thus this is somewhat other case than the case \cite{WJ}. But the effective Higgs mass $m/\sqrt\lambda$ incorporated naturally in the Minkowskian Higgs model with vacuum BPS monopoles quantized by Dirac, becomes zero in the limit of ``infinitely thick domain walls'' inside the appropriate discrete vacuum manifold $R_{YM}$,  (\ref {RYM}). It is just the ultraviolet region of the momentum space. On the other hand, in the limit of ``infinitely thin domain walls'' (it is just the infrared region of the momentum space), $m/\sqrt\lambda$ tends to a finite value \cite{disc} (the latter one can be treated as an  infrared cut-off).

Thus the  approach to the mass gap problem in the Minkowskian Higgs model quantized by Dirac involving  vacuum BPS monopole solution BPS monopole solutions and ``discrete vacuum geometry'' can be reduces to solving renorm-group equations \cite{Cheng} implicating the Wegner variable $m/\sqrt\lambda$,  that is a continuous function of the distance $r$.  Of course, these renorm-group equations would be in agreement with the first-order phase transition occurring therein \cite{disc}.

\bigskip
In general, in author's opinion, the Dirac fundamental quantization \cite{Dir} of gauge models seem to be having great perspectives in the future.

Really, the  FP heuristic quantization method \cite{FP1}, coming to fixing gauges in FP integrals, has  arisen at the end of 60-ies  of the past century, in despite of all its advantages at solving the problems associated with scattering processes in gauge theories, supplanted utterly from modern theoretical physics  the way of  references frames and initial and boundary conditions, the historically  arisen way in modern theoretical physics,  associated with Einstein (special and general) relativity \footnote{Recall  that Einstein (special and general) relativity is the historical and logical successor of the older Galilei  relativity theory and of Newton classical mechanics.}.  \par 
The FP heuristic quantization method \cite{FP1} retains in gauge theories only the realm of physical laws, bounded by the "absolutes", the S-invariants.  
But this   approach is fit, as we have discussed above, only for solving the problems associated with scattering processes in gauge theories, leaving "overboard" other  problems of modern theoretical physics,  including constructing bound states in QED and QCD. \par 
 In the present  study, with the example of the Minkowskian Higgs model quantized by Dirac \cite{Dir}, the author has attempted to attract the attention of the readers to the dramatic situation   that now arises in modern theoretical physics in connection with introducing the heuristic quantization method \cite{FP1} and  supplanting, by this method,  the  Dirac fundamental quantization  scheme \cite{Dir} (associated with the Hamiltonian approach to the quantization of gauge theories and attached to the definite reference frame).

\section{Acknowledges.}

  I am very grateful to Profs. R. Hofmann, Yu. P. Stepanovskij and V. I. Tkach
for the fruitful discussion and a series of useful remarks and Prof. V. Pervushin for
the help in the work on this article.
 
\enddocument